\documentclass[prc,twocolumn,showpacs,preprintnumbers,amsmath,amssymb,superscriptaddress,groupedaddress,nofootinbib,tightenlines]{revtex4}

\usepackage{graphicx}
\usepackage{amsmath}
\usepackage{bm}

\def\beqn{\begin{eqnarray}}
\def\eeqn{\end{eqnarray}}
\def\barr{\begin{array}}
\def\earr{\end{array}}
\def\btab{\begin{tabular}}
\def\etab{\end{tabular}}
\def\bite{\begin{itemize}}
\def\eite{\end{itemize}}
\def\bcen{\begin{center}}
\def\ecen{\end{center}}
\def\bwt{\begin{widetext}}
\def\ewt{\end{widetext}}
\def\qwp{Q_W^p}

\def\eq{\begin{equation}}
\def\ee{\end{equation}}
\def\nn{\nonumber}

\def\kdagger{K\hspace{-0.3cm}/\;}

\def\keldagger{k\hspace{-0.2cm}/}

\def\q2dagger{q_2\hspace{-0.35cm}/\;}

\def\sstw{\sin^2\theta_W}

\newcommand{\be}{\begin{equation}}
\newcommand{\bea}{\begin{eqnarray}}
\newcommand{\eea}{\end{eqnarray}}

\begin{document}
 {\preprint {NPAC-11-02}}
\title{Model-dependence of the  $\gamma Z$ dispersion correction to
  the parity-violating asymmetry in elastic $ep$ scattering}
\author{Mikhail Gorchtein} 
\affiliation{Center for Exploration of Energy and Matter, Physics Department,\\ 
Indiana University, Bloomington, IN 47403}
\email{mgorshte@indiana.edu}
\author{C. J. Horowitz} 
\affiliation{Center for Exploration of Energy and Matter, Physics Department,\\ 
Indiana University, Bloomington, IN 47403}
\author{Michael J. Ramsey-Musolf}
\affiliation{Department of Physics , University of Wisconsin-Madison, Madison, WI 53706 USA\\
and\\
Kellogg Radiation Laboratory, California Institute of Technology, Pasadena, CA 91125 USA}
\email{mjrm@physics.wisc.edu}
\date{\today}

\begin{abstract}
We analyze the dispersion correction to elastic parity violating 
electron-proton scattering due to $\gamma Z$ exchange. In particular, we 
explore the theoretical uncertainties associated with modeling
contributions of hadronic
intermediate states. Taking into account constraints from low- and high-energy, parity-conserving electroproduction
measurements, choosing different models for contributions from the non-resonant processes, and performing the corresponding flavor rotations to obtain the electroweak amplitude, we arrive at an estimate of the uncertainty in the
total contribution to the parity-violating asymmetry. At the kinematics of the Q-Weak experiment, we obtain a correction to the asymmetry equivalent to a shift in the proton weak charge of  
$(0.0054\pm0.0020)$. This should be compared to the value of the
proton's weak charge of $\qwp=0.0713\pm0.0008$ that includes Standard
Model contributions at tree level and one-loop radiative
corrections. Therefore, we obtain a new Standard Model prediction for
the parity-violating asymmetry in the kinematics of the Q-Weak
experiment of $(0.0767\pm0.0008\pm0.0020_{\gamma Z})$. 
The latter error leads to a relative uncertainty of 2.8\% in the
determination of the proton's weak charge, and is dominated by the uncertainty in the
isospin structure of the inclusive cross section. 
We argue that future parity-violating inelastic $ep$ asymmetry measurements at low-to-moderate $Q^2$ and $W^2$ could be exploited to reduce the uncertainty associated with the dispersion correction. Because the corresponding shift and error bar decrease monotonically with decreasing beam energy, a determination of the proton's weak charge with a lower-energy experiment or measurements of \lq\lq isotope ratios" in atomic parity-violation could provide a useful cross check on any implications for physics beyond the Standard Model derived from the Q-Weak measurement.
\end{abstract}
\maketitle

\section{Introduction}
Precise measurements of low-energy observables can provide powerful
probes of physics beyond the Standard Model that complement high
energy collider studies \cite{Erler:2004cx,RamseyMusolf:2006vr}. In particular, measurements of
parity-violating (PV) observables in atomic physics and electron
scattering have provided key tests of the neutral weak current sector
of the Standard Model and constrained possible new physics in this
sector\cite{erler,RamseyMusolf:2006vr,RamseyMusolf:1999qk,Young:2007zs}. In this work, we consider parity-violating (PV) elastic
scattering of longitudinally polarized electrons from hydrogen, which
is the subject of the Q-Weak experiment at the Jefferson Lab (JLab)
\cite{qweak}. This experiment draws on a rich history of
parity-violating electron scattering (PVES) at various facilities and
aims to provide the most precise determination of $\qwp$, the weak 
charge of the proton, ever made. 

In PVES, the weak charge is operationally defined through the forward scattering limit of the PV asymmetry:
\bea
\label{eq:apv}
A^{PV}&=&\frac{\sigma_+-\sigma_-}{\sigma_++\sigma_-}\ = \ 
\frac{G_Ft}{4\sqrt{2}\pi\alpha_{em}}\frac{W^\mathrm{PV}}{W^\mathrm{EM}}
\eea
where the ratio of response functions is defined below. Here and in
the rest of the article, $G_F$ denotes the Fermi constant, as taken
from the muon lifetime (often denoted by $G_\mu$).
The weak charge -- defined as a static property of the proton -- is
then the leading term the expansion of the ratio
${W^\mathrm{PV}}/{W^\mathrm{EM}}$ in powers of 
$t=-q^2$: 
\be
\label{eq:qweakdef}
\qwp=\lim_{t\to 0} \frac{W^\mathrm{PV}}{W^\mathrm{EM}}\Biggr\vert_{E=0}\ \ \ ,
\ee
where the reason for specifying zero beam energy $E$ will become apparent below.
In the one-boson exchange (OBE) approximation, the weak charge is just given by
\be
\qwp\Bigg\vert_\mathrm{OBE} =  -2\left[2C_{1u}+C_{1d}\right]\ \ \ ,
\ee
where the $C_{1q}$ characterize the effective four-fermion parity-violating electron-quark interaction
\bea
\label{ele-quark}
{\cal L} &=& \frac{G_F}{\sqrt{2}} \Big [\bar{e}\gamma^\mu \gamma_5 e
\big (C_{1u} \bar{u} \gamma_\mu u + C_{1d} \bar{d} \gamma_\mu d \big )
\nn\\
&&\;\;\;\;\;\;+\;\bar{e}\gamma^\mu e \big ( C_{2u} \bar{u} \gamma_\mu \gamma_5 u + C_{2d} \bar{d} \gamma_\mu \gamma_5 d \big ) \Big ]\,.
\eea

In the Standard Model, it is possible to make precise predictions for
the $C_{1q}$, including the effects of $\mathcal{O}(\alpha)$
electroweak radiative corrections \cite{erler,erler2,sirlin}. These
corrections include the effects of one-loop contributions to the gauge
boson and fermion propagators and gauge boson-fermion
vertices. Ultraviolet (UV)divergences are removed through
renormalization, and in what follows we will use the modified minimal
subtraction ($\overline{\mathrm{MS}}$ ) scheme for doing
so. 

Additional, UV-finite corrections arise from the two-boson
exchanges (\lq\lq box graphs"): $ZZ$, $W^+W^-$, $\gamma Z$ and
$\gamma\gamma$. Those involving two heavy vector bosons are dominated
by loop momenta of order $M_Z$ and are properly included in the
radiatively-corrected $C_{1q}$ coefficients. On the other hand, the
box graph corrections involving one or more photons are sensitive to
low-momentum scales where target-dependent hadronic structure effects
may be significant. In what follows, we focus on the $Z\gamma$ box
correction. For a review of recent work on the $\gamma\gamma$
corrections, see Ref. \cite{TPEreview}.

Recently, the $\gamma Z$ box graph contribution has been the subject of renewed scrutiny. In Refs. \cite{erler,erler2,Musolf:1990ts}, the short-distance part of this correction was computed, confirming the earlier computation of Ref. \cite{sirlin}. It carries a logarithmic dependence on the hadronic scale, $\Lambda_\mathrm{had}$, with the latter requiring the presence of a \lq\lq low energy constant"  $C_{\gamma Z}(\Lambda_\mathrm{had})$ to yield a result independent of the hadronic matching scale. The authors of Ref. \cite{erler,erler2} assigned a generous error to  $C_{\gamma Z}(\Lambda_\mathrm{had})$ associated with the difficult-to-compute long-distance hadronic effects. 

The authors of Ref. \cite{ja_chuck} subsequently observed that there
exists an additional contribution from the $\gamma Z$ box graph that
grows with the electron beam energy and that is independent of the
hadronic cutoff parameter\footnote{For related work considering the effects of the $\gamma Z$ box graph away from the forward limit -- relevant to the strange quark form factor determinations -- see Refs. \cite{Zhou:2007hr,Nagata:2008uv,Chen:2009mza}.}. Given the energy-dependence of this \lq\lq dispersion correction", it is more appropriate
to consider it as a new term in the PV asymmetry than as a contribution to the weak charge that is nominally a static property of the proton. Nevertheless, in the forward limit of Eq.~(\ref{eq:qweakdef}), its effect is to shift the apparent value of $\qwp$. Moreover, 
unlike the short-distance and $C_{\gamma
  Z}(\Lambda_\mathrm{had})$ terms that are suppressed by $1-4\sstw\approx
0.07$, the energy-dependent correction is not accidentally
suppressed. For the energy of the Q-Weak experiment, the authors of
Ref. \cite{ja_chuck} estimated that the correction was several
percent, raising the possibility that the estimated theoretical
uncertainty in the PV asymmetry could be larger than given in
Refs. \cite{erler,erler2}. 

A follow-up study \cite{sibirtsev} repeated the
computation of Refs. \cite{ja_chuck,deltagZ_conf} using a somewhat different
hadronic model framework and drawing upon recent structure function
measurements carried out at the Jefferson Laboratory. These authors
argued that the expressions used in Ref. \cite{ja_chuck} contained
numerical errors but nonetheless obtained a quantitatively similar
result for the size of the correction. An estimate of the uncertainty
in the correction was also provided, suggesting that the theoretical
uncertainty associated with the energy-dependent term is well below
the uncertainty quoted in Refs. \cite{erler,erler2}. Recently, another study
of this correction was reported in Ref.~\cite{carlson}. The latter
work employed yet another parametrization of virtual photoabsorption data
from Jefferson Lab, and a different treatment of the isospin structure
and of the uncertainty was applied. The results is consistent with that of Ref.~\cite{sibirtsev} with an
error bar that is also smaller than that of Refs. \cite{erler,erler2}.
We will review these works in
greater detail below. For the moment, we display in Table \ref{tab00} the results
of the previously mentioned studies along with the results of this work. While all of the recent
results (ours and Refs.~\cite{sibirtsev,carlson})  are consistent within quoted error bars, we obtain a larger
uncertainty by roughly a factor of two. As we discuss below, this larger theory uncertainty results from taking into
account hadronic model-dependence in computing the $\gamma Z$ dispersion correction.
\begin{table}[ht]
   \begin{tabular}{|l|l|l|l|}
\hline
Ref. \cite{ja_chuck} & Ref.
\cite{sibirtsev} & Ref. \cite{carlson} & This work \\
\hline
$(3\pm3)10^{-3}$ & $(4.7^{+1.1}_{-0.4}) 10^{-3}$
& $(5.7\pm0.9)10^{-3}$  & $(5.4\pm2.0)10^{-3}$\\
\hline
 \end{tabular}
\caption{Estimates for the dispersion correction Re$\,\Box_{\gamma Z}$
  obtained in various works, as indicated in the Table. Originally,
  Gorchtein and Horowitz in \cite{ja_chuck} only quoted the value of Re $\Box_{\gamma
    Z}\approx6$\%, as calculated relative to
  $1-4\sin^2\theta_W(0)\approx0.05$. This corresponds to the number
  given in Table.}
\label{tab00}
\end{table}

Obtaining a robust theoretical prediction for $A^{PV}$ in the Standard
Model is essential for the proper  interpretation of the asymmetry in
terms of possible contributions from physics beyond the Standard
Model. In light of the recent history and disagreements in the
literature on the question of the $\gamma Z$ box correction, we
revisit here the computations of
Refs. \cite{ja_chuck,deltagZ_conf,sibirtsev,carlson}. 
Our goal is three-fold. First, we seek to clarify the apparent disagreements about
the numerical factors in the analytic expressions for the
energy-dependent part of the $\gamma Z$ correction. Second, we attempt
to provide an estimate of the theoretical uncertainty associated with
hadronic modeling required for its computation. While the study of
Ref. \cite{sibirtsev} included an uncertainty associated with the
experimental data used as input for the calculation, no estimate of
the theoretical error related to the choice of model framework was
given. Finally, we discuss additional experimental input that would be
useful to improve the reliability of the calculated correction.

The remainder of our treatment of these points is organized as
follows. Section II outlines the elastic electron-nucleon scattering
kinematics and observables that are analyzed to one-loop order. 
In Section III, we derive a forward dispersion relation for the dispersion
corrections. In Section IV, we discuss the input in these sum rules,
perform an isospin decomposition of the inclusive electroproduction
data and isospin-rotate these data in order to obtain the inclusive
parity violating data. We combine different data sets to obtain an 
estimate of the uncertainty associated with such rotation in the
flavor space. Detailed discussion of the isospin rotation of the resonant
contributions is reported in Appendix A.
In Section V, we present our results for the dispersion
correction $\Box_{\gamma Z}$ and the respective theory uncertainty 
at the kinematics of the QWEAK experiment. 
Section VI is dedicated to the study of the $t$-dependence of
the dispersion correction that is important for translating the value
obtained from dispersion relation in the exact forward direction to
the experimental kinematics. In Section VII, we compare the
existing calculations of the energy-dependent dispersion $\gamma Z$
correction to the weak charge of the proton in detail. We close the
article with a short summary in Section VIII.

\section{PVES in the Forward Scattering Regime}
We consider elastic scattering of massless electrons off a nucleon, 
$e(k)+N(p)\to e(k')+N(p')$, in 
presence of parity violation (and in absence of $CP$-violation). The
scattering amplitude $T$ can be cast in the following form involving 
six scalar amplitudes $f_i(\nu,t)$, $i=1,2,...,6$,
\bwt
\beqn
T&=&\frac{4\pi\alpha_{em}}{-t}\bar{u}(k')\gamma_\mu u(k)
\bar{N}(p')\left[f_1\gamma^\mu+f_2i\sigma^{\mu\alpha}
\frac{\Delta_\alpha}{2M}+f_3\frac{P^\mu\kdagger}{M^2}\right]N(p)\nn\\
&-&\frac{G_F}{2\sqrt{2}}\bar{u}(k')\gamma_\mu\gamma_5 u(k)
\bar{N}(p')\left[f_4\gamma^\mu+f_5i\sigma^{\mu\alpha}
\frac{\Delta_\alpha}{2M}\right]N(p)
\;-\;\frac{G_F}{2\sqrt{2}}f_6\,\bar{u}(k')\gamma_\mu u(k)
\bar{N}(p')\gamma^\mu\gamma_5N(p)\label{eq:ampl},
\eeqn
\ewt
where only electromagnetic and weak neutral currents are considered. 
$G_F$ stands for the Fermi constant, as taken from the muon
lifetime, according to the $\overline{\mathrm{MS}}$ scheme. 
The amplitudes $f_{1,2,3}$ are parity conserving (PC), and $f_{4,5,6}$ are 
explicitly parity violating (PV). Above, $k(k')$ stands for the initial (final)
electron momenta, and $p(p')$ for the initial (final) nucleon momenta,
respectively, and $M$ denotes the mass of the nucleon (we take
$M_n\approx M_p\equiv M$). All six amplitudes are functions of energy
$\nu=\frac{PK}{M}$ (with $K=\frac{k+k'}{2}$ and $P=\frac{p+p'}{2}$) 
and the elastic momentum transfer is $t=\Delta^2<0$, with $\Delta=k-k'=p'-p$.
At tree level (one boson exchange, OBE) and to leading order in $G_F$
and $\alpha_{em}$, 
the amplitudes $f_i$ reduce to the electromagnetic and weak form
factors of the nucleon (the index $N$ takes values $p,n$ denoting proton
and neutron, respectively), 
\beqn
f_1^{N,OBE}(\nu,t)&=&{\cal F}_1^{\gamma N}(t) \nn\\
f_2^{N,OBE}(\nu,t)&=&{\cal  F}_2^{\gamma N}(t)\nn\\
f_3^{N,OBE}(\nu,t)&=&0\nn\\
f_4^{N,OBE}(\nu,t)&=&g_A^e{\cal F}_1^{Z N}(t)\nn\\
f_5^{N,OBE}(\nu,t)&=&g_A^e{\cal F}_2^{Z N}(t)\nn\\
f_6^{N,OBE}(\nu,t)&=&g_V^eG_{A,N}^{e}(t)\,.
\eeqn
\indent
Above, $g_V^e=-(1-4\sin^2\theta_W)$ and $g_A^e=1$. 
Radiative corrections induce terms $\delta f_i\sim\alpha_{em}$, 
leading generically to $f_i=f_i^{OBE}(t)+\delta f_i(\nu,t)$. 
We denote the usual Dirac (Pauli) form factors by ${\cal F}_{1,2}^{\gamma
  N}$, respectively, and the nucleon axial form factor at tree level
by $G_{A,N}^e$. Similarly, ${\cal F}_{1,2}^{Z N}$ stand for the form factors
describing the vector coupling of the $Z$ to the nucleon. 
One introduces the conventional combinations, 
\beqn
 G_M^{\gamma }&=&f_1^N+f_2^N\,, \nn\\
 G_E^{\gamma }&=&f_1^N-\tau f_2^N\,,\nn\\
 G_M^{Z}&=&\frac{1}{g_A^e}(f_4^N+f_5^N)\,,\nn\\ 
 G_E^{Z}&=&\frac{1}{g_A^e}(f_4^N-\tau f_5^N), 
\eeqn
with $\tau=\frac{-t}{4M^2}$.  In absence of radiative corrections, these
 amplitudes reduce to the electroweak Sachs form factors 
$G_{E,M}^{\gamma ,\,Z}$.
In terms of these generalized form factors, the
 unpolarized cross section on a nucleon target $N$ can be written as
\beqn
\frac{d\sigma^N}{d\Omega_{Lab}}&=&\frac{4\alpha_{em}^2\cos^2\frac{\theta}{2}}{t^2}\frac{{E'}^3}{E}
\frac{\tau\sigma_R^N}{\varepsilon(1+\tau)},
\eeqn
with $\theta$ the electron Lab scattering angle, $E(E')$ the incoming
(outgoing) electron Lab energy, and $\varepsilon=(1+2(1+\tau)\tan^2\frac{\theta}{2})^{-1}$
 the virtual photon longitudinal polarization parameter. The reduced
 cross  section $\sigma_R^N$, up to and including terms of order $\alpha_{em}$, reads
\beqn
\sigma_R^N&=&| G_M^{\gamma }|^2 + \frac{\varepsilon}{\tau}
| G_E^{\gamma }|^2 +
2\varepsilon\frac{\nu}{M}(G_M^{\gamma }+\frac{1}{\tau}G_E^{\gamma })
{\rm Re}f_3^N\nn\\
\label{eq:sigmaR}
\eeqn
\indent
In what follows, we will concentrate on the case of electron-proton
scattering. Therefore, we will understand $N=p$ everywhere and
suppress the index $N$ in all expressions, unless explicitly stated
otherwise.

The parity violating asymmetry is defined in Eq.~(\ref{eq:apv})
with the ratio of the response functions is given by 
\bwt
\bea
\frac{W^\mathrm{PV}}{W^\mathrm{EM}}&=&{\rm Re}
\frac{ G_M^\gamma G_M^{Z*}+\frac{\varepsilon}{\tau}
 G_E^\gamma G_E^{Z*} + \varepsilon\frac{\nu}{M}f_3
( G_M^{Z*}+\frac{1}{\tau} G_E^{Z*})
+\frac{\varepsilon'}{\tau} G_M^\gamma f_6^*}{\sigma_R}\,.\label{eq:apv1}
\eea
\ewt
\indent
Here, $\sigma_{\pm}$ are the cross sections for positive and negative helicity electrons,
 and $\varepsilon'=\sqrt{\tau(1+\tau)(1-\varepsilon^2)}$.

Since we are interested in very forward scattering angles 
$\theta\approx8^\circ$ corresponding to the Q-Weak kinematics \cite{qweak}, 
thus $\tau<10^{-3}$, the expressions for the cross section and PV
asymmetry can be fruther simplified.

For the reduced cross section the leading contribution in
Eq. (\ref{eq:sigmaR}) comes from the $G_E^2$ term, and we obtain
\beqn
\sigma_R&=&\frac{1}{\tau}({\cal F}_1^\gamma)^2 (1+\tau\delta_{kin}^{\sigma}
+2{\rm Re}\bar\delta_{RC}^{\sigma})+2{\rm Re}\Box_{\gamma\gamma}.
\eeqn
\indent
The three distinct corrections quoted above are defined
as follows: $\delta_{kin}^{\sigma}$ is a kinematic correction that
arises at tree level due to the magnetic part and other 
subleading kinematic effects of order $\tau,(1-\epsilon)$, 
that do not contain ${\cal O}(\alpha_{em})$ effects;
$\bar\delta_{RC}^{\sigma}$ stands for order ${\cal O}(\alpha_{em})$
corrections that are energy-independent (such as 
vacuum polarization, self energy and vertex corrections); 
finally, $\Box_{\gamma\gamma}$ denotes the two-photon exchange
correction that is an energy-dependent ${\cal O}(\alpha_{em})$ 
correction.

Similarly, for the PV asymmetry the leading order contribution
in Eq. (\ref{eq:apv1}) originates from the $G_E^{Z*} G_E^\gamma$ term. 

As discussed in Ref.~ \cite{erler}, the Standard Model prediction for the PV asymmetry in the forward regime can be expressed as 
\bwt
\beqn
A^{PV}&=&\frac{G_Ft}{4\sqrt{2}\pi\alpha_\mathrm{em}}
\left[(1+\Delta\rho+\Delta_e)(1-4\sin^2\hat\theta_W(0)+\Delta_e')+\Box_{WW}+\Box_{ZZ}+\Box_{\gamma
Z}\right]+\cdots,
\eeqn
\ewt
where $\hat\theta_W(0)$ is the running weak mixing angle in the $\overline{MS}$ scheme at zero momentum transfer\cite{erler2}.
The correction $\Delta\rho$ is a universal radiative correction to the relative normalization of the neutral and charged current amplitudes; the $\Delta_e$ and $\Delta_e^\prime$ give, respectively, non-universal corrections to the axial vector $Zee$ and $\gamma ee$ couplings; the $\Box_{VV}$ for $V=W,Z,\gamma$ give the non-universal box graph corrections; and the \lq\lq $+\cdots$" indicate terms that vanish  with higher powers of $t$ in the forward limit, such as those arising from the magnetic and strange quark form factors and the two-photon dispersion correction, $\Box_{\gamma\gamma}$. The weak charge of the proton, considered as a static property, is given by the quantity in the squark brackets in the zero-energy limit. 

Within the radiative corrections, the TBE effects are separated
explicitly. This is done because the TBE corrections, unlike other
corrections in the above equation, are in general $\nu$ and $t$-dependent. 
In particular, the $\nu$ (or $\varepsilon$) dependence of the 
$\gamma\gamma$-box is believed to be responsible for the discrepancy 
between the Rosenbluth and polarization transfer data for $G_E^\gamma/G_M^\gamma$
\cite{gegm_exp}. It should be noted that in the exact
forward direction $\Box_{\gamma\gamma}$ vanishes as a consequence of
electromagnetic gauge invariance. 

The $WW$ and $ZZ$-box diagrams were first
considered in \cite{sirlin}, and subsequently investigated in Refs.~\cite{mjrm,erler}.  The contribution from $\Box_{WW}$ in particular is relatively large.  
Both corrections are
$\nu$-independent at any hadronic energy scale since they
are dominated by exchange of hard momenta in the loop $\sim M_W,M_Z$. Higher-order perturbative
QCD corrections to $\Box_{WW}$ and $\Box_{ZZ}$ were computed in Ref.\cite{erler}, and the overall theoretical uncertainty
associated with these contributions is well below the expected uncertainty of the QWEAK experiment.

In contrast to $\Box_{WW}$ and $\Box_{ZZ}$, $\Box_{\gamma Z}$ receives substantial contributions from loop momenta at all scales. For the electron energy-independent contribution, this situation leads to the presence of a large logarithm $\ln M_Z/\Lambda_\mathrm{had}$ where $\Lambda_\mathrm{had}$ is a typical hadronic scale\cite{sirlin,mjrm,erler}. Since the asymmetry must be independent of the latter, $\Box_{\gamma Z}$ includes also a \lq\lq low-energy constant" $C_{\gamma Z}(\Lambda_\mathrm{had})$ whose hadronic scale dependence compensates for that appearing in the logarithm. An analogous $W\gamma$ box correction enters the vector current contribution to neutron and nuclear $\beta$-decay. Importantly for the PV asymmetry, these energy-independent $\gamma Z$ box contributions are suppressed by $1-4\sin^2\theta_W$, thereby suppressing the associated theoretical uncertainty. 


In Ref.~\cite{ja_chuck},  the $\gamma Z$-box contribution was re-examined in the framework 
of dispersion relations and it was found that it possesses a
considerable energy dependence, so that at energies in the GeV
range its value can differ significantly from that found at zero
energy. Moreover, the energy-dependent correction contains a term that is not $1-4\sin^2\theta_W$ suppressed, so the theoretical uncertainty associated with hadronic-scale contributions is potentially more significant. 
This energy dependence comes through contributions from
hadronic energy range inside the loop that cannot be calculated
reliably using perturbative techniques. 

At present, a complete first principles computation is not feasible, forcing one to rely on hadronic modeling. For a proper interpretation of the PV asymmetry, it is thus important to investigate the theoretical hadronic model uncertainty. The remainder of the paper is devoted to this task. In so doing, we will attempt to reduce this model uncertainty by relating -- wherever possible -- contributions from hadronic intermediate states to experimental parity-conserving electroproduction data through the use of a dispersion relation and isospin rotation. As a corollary, we will also identify future experimental measurements, such as those of the parity-violating inelastic asymmetry in the regime of moderate $Q^2$ and $W$, that could be helpful in reducing the theoretical uncertainty.


\section{Dispersion corrections}
\begin{figure}[h]
\includegraphics[width=2in]{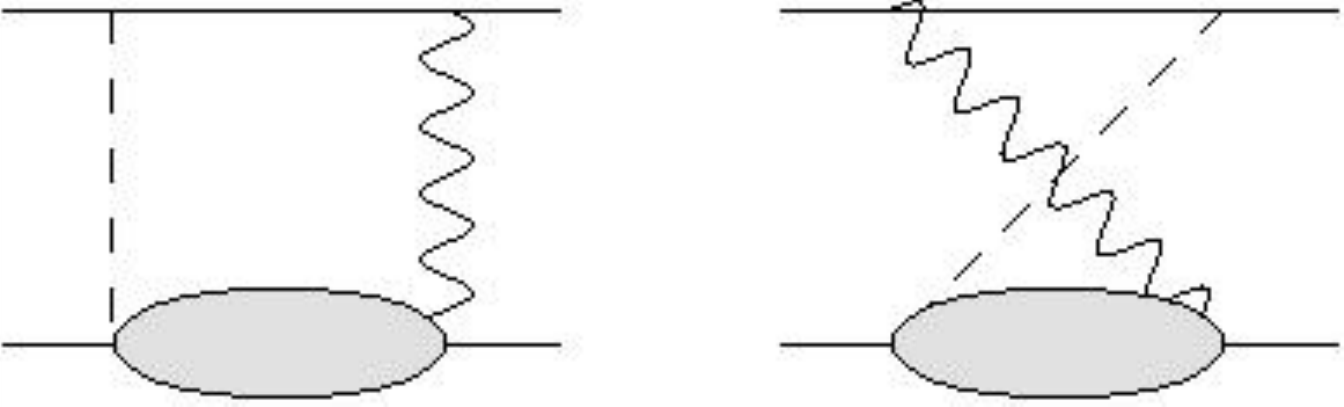}
\caption{Direct and crossed diagrams for $\gamma Z$-exchcange. Dashed
  lines correspond to an exchange of a $Z$-boson, and wavy lines to
  an exchange of a photon. The blob stands for an inclusive sum over
  intermediate hadronic states.} 
\label{fig:box}
\end{figure}

To calculate the real part of the $\gamma Z$ direct and crossed box
diagrams showed in Fig. \ref{fig:box}, 
we follow \cite{ja_chuck} and adopt a dispersion relation formalism. We start with the 
calculation of the imaginary part of the direct box (the crossed 
box contribution to the real part will be calculated using crossing), 
\beqn
{\rm Im}T_{\gamma Z}
\,=\,-\frac{G_F}{\sqrt{2}}\frac{e^2}{(2\pi)^3}\int\frac{d^3\vec{k}_1}{2E_1}
\frac{l_{\mu\nu}\cdot W_{\gamma Z}^{\mu\nu}}{Q^2(1+Q^2/M_Z^2)},
\label{eq:impart}
\eeqn
where $Q^2=-(k-k_1)^2$ 
denotes the virtuality of the exchanged photon and $Z$ (in the 
forward direction they carry exactly the same $Q^2$), and 
we explicitly set the intermediate electron on-shell. In the center of
mass of the (initial) electron and proton, one has 
$E_1=\frac{s-W^2}{2\sqrt{s}}$, with $s$ the 
full c.m. energy squared and $W$ the invariant mass of the intermediate 
hadronic state. Note that for on-shell intermediate states, the exchanged 
bosons are always spacelike. 

The leptonic tensor is given by
\beqn
l_{\mu\nu}=\bar{u}(k')\gamma_\nu\keldagger_1\gamma_\mu
(g_V^e+g_A^e\gamma_5) u(k).
\eeqn
\indent
We next turn to the lower part of the diagrams in Fig. \ref{fig:box}. 
The blobs stand for an inclusive sum over all possible hadronic
intermediate states, starting from the ground state (i.e., the nucleon
itself) and on to a sum over the whole nucleon photoabsorption spectrum. 
The case of the elastic hadronic intermediate state was considered in
\cite{blunden}. Here, we concentrate on the inelastic contribution. 
Such contributions arise from the absorption of a photon (weak
boson). 
In electrodynamics, for a given material, the
relation between its refraction coefficient and the dependence of the
latter on the photon frequency
(i.e., dispersion) on one hand, and the photoabsorption spectrum of
that material on the other hand, 
is historically called a dispersion relation. It is exactly
this dependence of the forward scattering amplitude $f_4(\nu,0)$ (see
Eq. (\ref{eq:ampl})) on the energy that arises from its relation to
the electroweak $\gamma(Z)$-absorption spectrum that is the scope 
of an investigation in this work. This
explains the origin of the term ``dispersion correction'' used for the
inelastic contributions to the $\gamma Z$-box correction. 

In the forward direction, the imaginary part of the 
doubly virtual ``Compton scattering'' ($\gamma^*p\to Z^*p$) amplitude is 
given in terms of the interference structure functions $F^{\gamma Z}_{1,2,3}(x,Q^2)$, with 
$x=\frac{Q^2}{2Pq}$ the Bjorken variable. Making use 
of gauge invariance of the leptonic tensor, we have
\beqn
\frac{1}{2\pi}W_{\gamma Z}^{\mu\nu}=-g^{\mu\nu}F^{\gamma Z}_1 
+ \frac{P^\mu P^\nu}{Pq}F^{\gamma Z}_2
+ i\epsilon^{\mu\nu\alpha\beta}\frac{P_\alpha q_\beta}{Pq}F^{\gamma Z}_3\nn\\
\label{eq:hadron_tensor}
\eeqn
\indent
Contracting the two tensors, one obtains after a little algebra two
contributions that are due respectively to the axial and vector couplings of the
$Z$ to the electron,
\begin{widetext}
\beqn
{\rm Im}\Box_{\gamma {Z_A}}(\nu)&=&\alpha_\mathrm{em} g_A^e
\int_{W^2_\pi}^s\frac{dW^2}{(s-M^2)^2}
\int_0^{Q^2_{max}}\frac{dQ^2}{1+\frac{Q^2}{M_Z^2}}
\left[F^{\gamma Z}_1
+ \frac{s(Q^2_{max}-Q^2)}{Q^2(W^2-M^2+Q^2)}F^{\gamma Z}_2\right]\nn\\
{\rm Im}\Box_{\gamma {Z_V}}(\nu)&=&-\alpha_\mathrm{em} g_V^e
\int_{W^2_\pi}^s\frac{dW^2}{(s-M^2)^2}
\int_0^{Q^2_{max}}\frac{dQ^2}{1+\frac{Q^2}{M_Z^2}}
\left(\frac{2(s-M^2)}{W^2-M^2+Q^2}-1\right) F^{\gamma Z}_3,
\label{imdeltagz}
\eeqn
\end{widetext}
where the imaginary parts $\mathrm{Im}\Box$ will appear in a dispersion relation for the real parts in Eq.~(\ref{eq:dr}) below. The full correction  is the sum of the two, 
\beqn
{\rm Im}\Box_{\gamma {Z}}(\nu)&=&{\rm Im}\Box_{\gamma {Z_A}}(\nu)
+{\rm Im}\Box_{\gamma {Z_V}}(\nu).
\eeqn
\indent
In Eqs.~(\ref{imdeltagz}),  $W^2_\pi=(M+m_\pi)^2$ stands for the pion production threshold, and 
the $Q^2$-integration is constrained below a maximum value
\be
\label{eq:q2max}
Q^2_{max}=\frac{(s-M^2)(s-W^2)}{s}
\ee
as a condition of on-shell intermediate 
states for an imaginary part calculation.
Eq. (\ref{imdeltagz}) is in agreement with Refs. \cite{sibirtsev,
  carlson}. In particular, we confirm the correctness of the claim made in
Ref. \cite{sibirtsev} that in Ref. \cite{ja_chuck} a factor of 2 was
missing. 

In order to write down the dispersion relation for the function 
$\Box_{\gamma{Z}}(\nu)$, 
one should consider its behavior under crossing. 
We distinguish two contributions, $\Box_{\gamma {Z_V}}$ and 
$\Box_{\gamma {Z_A}}$ that have different crossing behavior \cite{ja_chuck}:
\beqn
\Box_{\gamma {Z_A}}(-\nu)&=&-\Box_{\gamma {Z_A}}(\nu)\nn\\
\Box_{\gamma {Z_V}}(-\nu)&=&+\Box_{\gamma {Z_V}}(\nu)
\eeqn
Correspondingly, the two contributions obey dispersion relations of two different
forms,
\beqn
{\rm Re}\Box_{\gamma {Z_A}}(\nu)&=&\frac{2\nu}{\pi}\int_{\nu_\pi}^\infty
\frac{d\nu'}{\nu'^2-\nu^2}{\rm Im}\Box_{\gamma{Z_A}}(\nu')\nn\\
{\rm Re}\Box_{\gamma {Z_V}}(\nu)&=&\frac{2}{\pi}\int_{\nu_\pi}^\infty
\frac{\nu'd\nu'}{\nu'^2-\nu^2}{\rm Im}\Box_{\gamma{Z_V}}(\nu')
\label{eq:dr}
\eeqn
\indent
where the presence or absence of the factor of $\nu^\prime$ in the integrands follows from the behavior of the $\mathrm{Im}\Box$ under crossing symmetry.

The result in Eq. (\ref{eq:dr}) gives a model-independent relation
between the dispersion correction to the weak charge of the proton and 
the parity violating structure functions appearing in Eq.~(\ref{imdeltagz}). This relation does
not rely on any assumption, other than the neglect of 
higher order radiative corrections and the number of subtractions needed for convergence of the dispersion relation.
The advantage for this formulation is that the $F_k^{\gamma Z}$ are in principle measurable. 
However, in absence of any detailed 
parity violating inclusive electron scattering data, the input in the
dispersion integral will depend on a model. 
In the following, we will investigate the extent to which this model dependence can 
be constrained by existing or future experimental data.

\section{Input to the dispersion integral}

In the previous section, the contribution of the forward hadronic tensor to the 
box diagram was considered. In this section, we will address the possibility of 
relating the interference hadronic tensor of Eq. (\ref{eq:hadron_tensor})
\beqn
W^{\mu\nu}_{\gamma Z}=\frac{1}{2}\int d^4ze^{iqz}\langle
N|T\left[J^\nu_{em}(z)J^\mu_{NC_V}(0)\right]|N\rangle
\eeqn
to the pure electromagnetic one, 
\beqn
W^{\mu\nu}_{\gamma \gamma}=\frac{1}{2}\int d^4ze^{iqz}\langle
N|T\left[J^\nu_{em}(z)J^\mu_{em}(0)\right]|N\rangle.
\eeqn

Using unitarity, we rewrite these matrix elements as an inclusive sum over 
intermediate hadronic states,
\beqn
{\rm Im}W^{\mu\nu}_{\gamma Z}
&\!=\!&\frac{1}{2}\!\int \!\!d^4ze^{iqz}\!\!\left[
\sum_X\langle N|J^\nu_{em}(z)|X\rangle\langle
X|J^\mu_{NC_V}(0)|N\rangle\right.\nn\\
&&+\;\left.\sum_X\langle N|J^\nu_{NC_V}(z)|X\rangle\langle
X|J^\mu_{em}(0)|N\rangle\right]\label{eq:Wgz_def}
\eeqn
and 
\beqn
{\rm Im}W^{\mu\nu}_{\gamma \gamma}
&=&\frac{1}{2}\int d^4ze^{iqz}\sum_X\langle N|J^\nu_{em}(z)|X\rangle\langle X|J^\mu_{em}(0)|N\rangle\nn\\
\label{eq:Wgg_def}
\eeqn
respectively. We now proceed to investigate the possible relationships between the products of transition matrix elements appearing in each inclusive sum (\ref{eq:Wgz_def}) and (\ref{eq:Wgg_def}). 

\begin{figure}
\includegraphics[width=8cm]{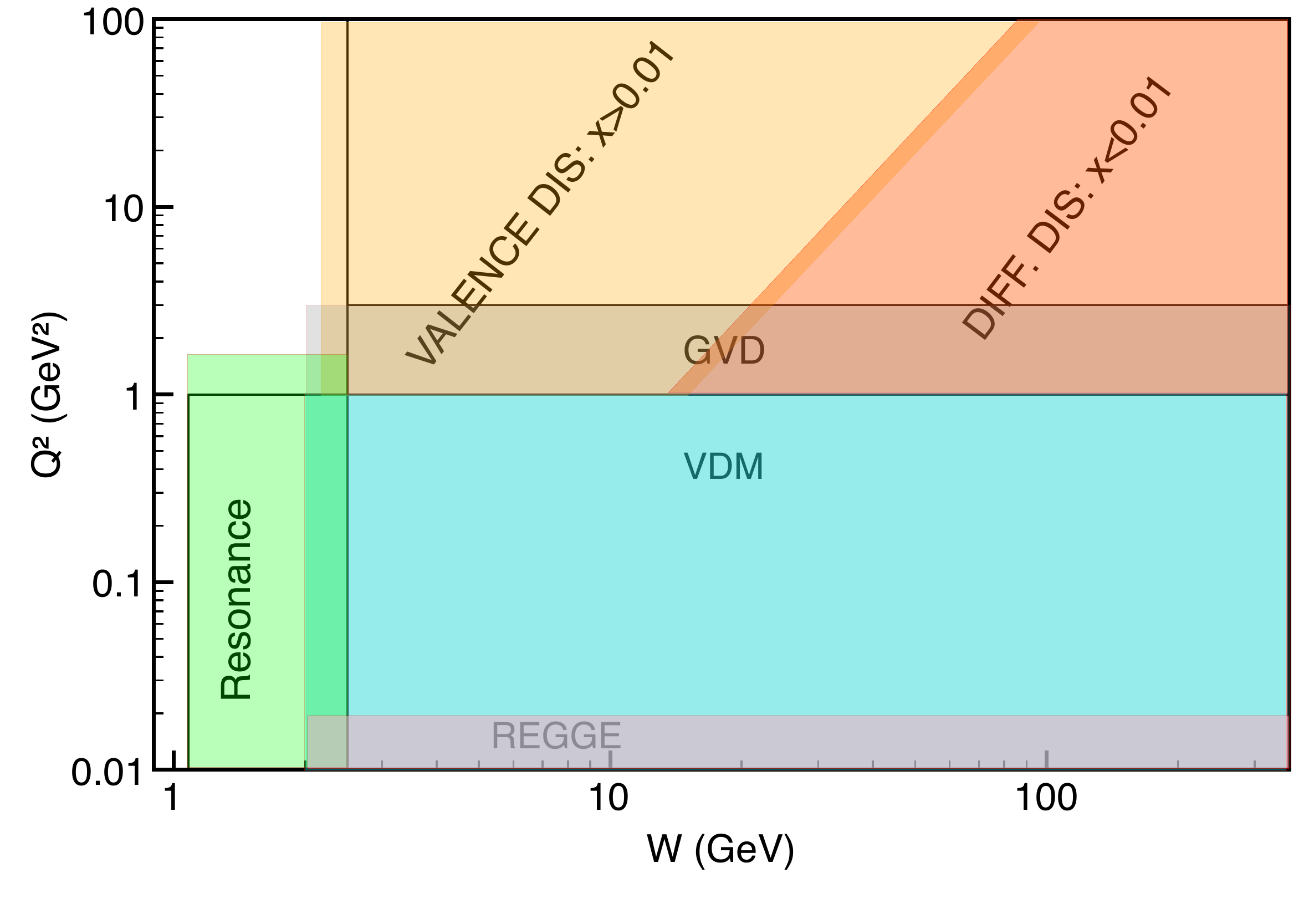}
\caption{The plane $Q^2$ vs. $W$ and kinematic regions
  corresponding 
roughly to various physical contributions.} 
\label{fig:WQdiagram}
\end{figure}

Theoretically, calculating the full set of contributions to the inclusive sum  represents 
a fundamental difficulty since in QCD, the basis for intermediate states $X$ 
is infinite, and the matrix elements are non-perturbative. Under certain kinematic conditions, 
one can organize this basis into leading and subleading (kinematically 
suppressed) sub-sets. We depict this situation schematically in
Fig. \ref{fig:WQdiagram}, where we show in the $Q^2-W^2$ plane the
approximate kinematic areas where various mechanisms dominate. 
At high energy and $Q^2$, and finite Bjorken $x$, the 
leading set of states is $X=q+X'$ ($q$ denotes a quark), where to leading order in $1/Q$, $X'$ is a 
spectator. Thus, in this regime the electromagnetic (weak) current
directly probes a single quark within the 
nucleon, and gives access to the parton distribution functions (deep
inelastic scattering, DIS in Fig. \ref{fig:WQdiagram}).
At high energy and $Q^2$, and small $x$, however, the picture changes, 
as the leading set is 
$X=\bar{q}q+N$. In this regime, the photon polarizes the QCD vacuum at the 
periphery of the hadron, and the resulting $\bar{q}q$-pair forms a color dipole 
that interacts with the nucleon (diffractive DIS in
Fig. \ref{fig:WQdiagram}).
This picture was first realized in the 
Vector Meson Dominance model (VDM) that capitalized on the fact that since vector 
mesons and the photon have the same quantum numbers, the latter can fluctuate 
into former  \cite{vdm,gvd}. 
This simple model works quite well at
low $Q^2$ (VDM area in Fig. \ref{fig:WQdiagram}). Such ``hadron-like''
behavior of a photon in scattering processes also results in the e.-m. data
following the Regge behavior, as a function of $W$ (respective Regge area
in Fig. \ref{fig:WQdiagram}).
At higher values of $Q^2$, rescattering effects in vector
meson-nucleon scattering become increasingly important but can still
be accounted for in what is called the \lq\lq generalized VDM" (GVDM region in
Fig. \ref{fig:WQdiagram}).  
At low energies, the relevant degrees of freedom are hadronic (that is, highly 
non-perturbative), $X=N,\pi N, \pi\pi N, N^*, \Delta$ etc. In this regime, the inelastic 
cross section is tyically dominated by resonances on top of a non-resonant background (Resonance area in
Fig. \ref{fig:WQdiagram}).  
The boundaries of each kinematic region are, of course, approximate.
Their meaning is that the farther one departs from a kinematical
region, to the lesser extent the respective mechanism works. 
Consequently, a large area on the $W^2-Q^2$ plane, that overlaps with
all the depicted regions but not covering them completely is the
so-called shadow region where none of the mechanisms can be considered
as fully dominant. 

If data for the $\gamma Z$ interference cross section existed throughout all these 
distinct regimes, we would not need to know details of any of the 
aforementioned models. In principle, such data could be obtained with measurements of
the PV inelastic asymmetries in the various kinematic regimes shown in Fig. \ref{fig:WQdiagram}.
At present, however, either no or very poor data on PV inelastic scattering exist. 
Consequently, we will instead pursue an alternate strategy, endeavoring to make use of extensive data sets for real and virtual photoabsorption that exist through vast kinematic region in energy and $Q^2$. 
To that end, we will rely on models that adequately describe the photoabsorption cross section in different regimes and for each attempt to establish relationships between the matrix elements 
$\langle X|J^\mu_{em}|N\rangle$ and $\langle X|J^\mu_{NC}|N\rangle$
for each intermediate hadronic state $|X\rangle$ of definite
isospin. We will approach this problem by extracting the
electromagnetic matrix elements from inclusive e.-m. data, 
and then isospin-rotate every such matrix element. We begin with a brief review of the experimental situation and discuss various model descriptions.

\subsection{Real and virtual photoabsorption data}
We find that the dispersion integral for
Re$\Box_{\gamma Z}$ is dominated by moderate values of $W\lesssim5$
GeV and $Q^2\lesssim3$ GeV$^2$ (see Fig. \ref{fig:kinreg} in Section V). Consequently, we need to analyze in
detail contributions from the resonance regime and portions of what we
have called the VDM, GVDM, and regge regimes. 
Our goal will be to draw upon existing experimental data for inclusive
and semi-inclusive electromagnetic data to infer the $\gamma Z$
interference structure functions that appear in the dispersion
integrals. To that end, we first summarize the experimental situation.
\begin{itemize}
\item{} Real photoabsorption cross sections have been measured from the pion threshold to very
high energies \cite{bloom69,armstrong72,caldwell73,caldwell78,chekanov02,vereshkov03}
\item{} Virtual photoabsorption data: 
high precision data from the JLab E94-110
\cite{E94} and the preliminary data from the E00-002 \cite{E00} experiments 
are available in the resonance region; 
in the DIS region, we quote the data for the DIS structure function
$F_2$ from SLAC NMC Collaboration \cite{F2_NMC}, FNAL E665
collaborations \cite{E665} and DESY H1 Collaboration \cite{F2_H1}. 
\end{itemize}
While it is equally possible to use structure functions to describe
resonance data, in the following we opt to 
use total photoabsorption cross sections
with transverse or longitudinal (for
virtual photons only) photon polarization. These cross sections are
unambiguously related to the electromagnetic structure functions,
\beqn
&&\sigma^{\gamma p}_T(W^2,Q^2)=\frac{8\pi^2\alpha}{W^2-M^2}F^{\gamma\gamma}_1(x,Q^2) 
\label{eq:f12-sigmaTL}\\
&&\sigma^{\gamma p}_L(W^2,Q^2)=\frac{8\pi^2\alpha}{W^2-M^2}\nn\\
&&\left[
  \left(\frac{1}{2x}+\frac{2M^2}{W^2-M^2+Q^2}\right)F^{\gamma\gamma}_2(x,Q^2)
-F^{\gamma\gamma}_1(x,Q^2)\right]\nn
\eeqn
with the usual Bjorken scaling variable $x=\frac{Q^2}{W^2-M^2+Q^2}$.
This choice is convenient because in what follows, we will address
transitions between helicity states of the nucleon and resonances, and
it is preferrable to work with matrix elements of the electromagnetic
current with definite helicities. As is evident from
Eq. (\ref{eq:f12-sigmaTL}), the two helicity states are mixed in
$F_2$. 
Similar relations hold between the interference cross sections 
$\sigma^{\gamma Z,p}_{T,L}(W^2,Q^2)$ and interference structure
functions $F^{\gamma Z}_{1,2}(x,Q^2)$. Note that the definition of the
transverse and longitudinal polarizations of the photon and the
$Z$-boson are identical since in both cases they are fixed by the 
lepton kinematics of the reaction $e+p\to e'+X$. 

Real photoabsorption data exhibit the following general features: i) a resonance
structure on top of ii) a smooth non-resonant background 
between the threshold of pion production and $W\sim2-2.5$
GeV, and iii) Regge behavior at high values of $W$ with the cross
section that grows slowly with energy, $\sigma^{\gamma p}_\mathrm{tot}\sim
(W^2)^{\alpha_P-1}$, with $\alpha_P\sim1.095$ the parameter of the
pomeron.
\beqn
\sigma_\mathrm{tot}^{\gamma p}(W^2)&=&\sigma_\mathrm{Res}^{\gamma p}(W^2)
+\sigma_\mathrm{Bkgd}^{\gamma p}(W^2),
\nn\\
\sigma_\mathrm{tot}^{\gamma
  p}(W^2\to\infty)&\rightarrow&\sigma_\mathrm{Regge}^{\gamma p}(W^2)\sim (W^2)^{\alpha-1},
\eeqn
where $\alpha=\alpha_P,\alpha_{f_2}$, etc. stand for pomeron and Regge
trajectories. 
In this work, the most recent fit in terms of two trajectories
(pomeron plus $f_2$) is used \cite{regge_fit}
\beqn
\sigma^{\gamma p}_\mathrm{Bkgd}&=&f_\mathrm{thr}
\left[(145.0\pm2.0)\,\mu b \left(\frac{W^2}{W_0^2}\right)^{-0.5}\right.\nn\\
&+&\left.(63.5\pm0.9)\,\mu b \left(\frac{W^2}{W_0^2}\right)^\epsilon\right]
\eeqn
with parameter of the pomeron $\epsilon={0.097\pm0.002}$.
The threshold factor $f_\mathrm{thr}$ is necessary to make the continuation of
the Regge fit into the resonance region meaningful. In this work, we
take it in the same form as in \cite{bianchi}
\beqn
f_\mathrm{thr}=1-\exp\left[-\frac{W^2-(M+m_\pi)^2}{M^2}\right].
\eeqn
\indent

For virtual photons in the range of $W^2,Q^2$ of interest here, the
picture remains the same, with the $Q^2$-dependence of the resonance
contributions described by the form factors measured for a number of
resonances, at least in certain channels.
\begin{figure}[h]
\includegraphics[width=8.6cm]{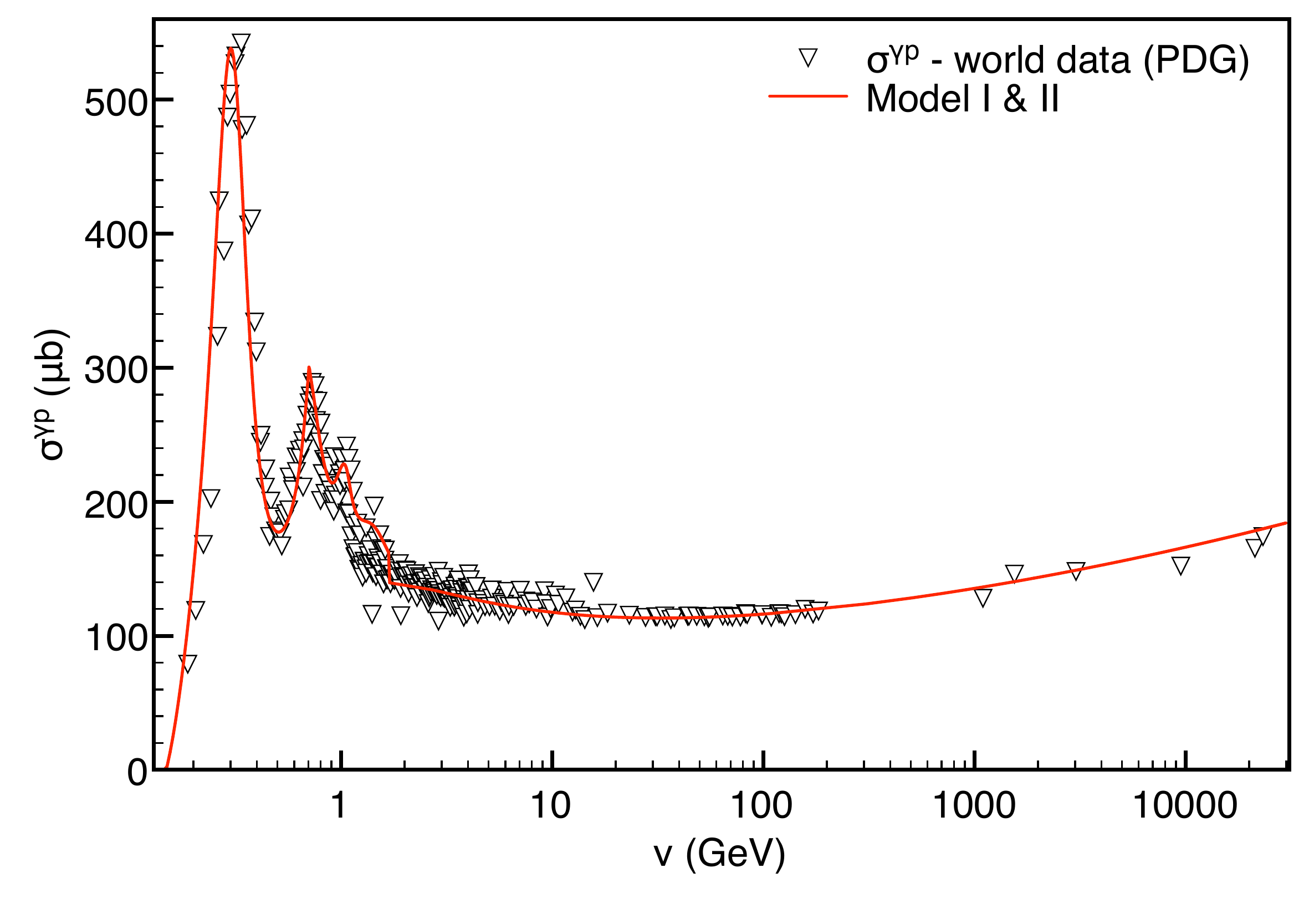}
\caption{(Color online) World data on total photoabsorption 
\cite{bloom69,armstrong72,caldwell73,caldwell78,chekanov02,vereshkov03}
(see \cite{pdg} for the complete list) compared to
  the two models described in the text. The experimental errors are
  not shown.} 
\label{fig:sigmatot}
\end{figure}
We will next specify two models that provide a smooth extrapolation between
the real photoabsorption data and the virtual photoabsorption data and
that can to certain extent be used to describe data all the way up
into the diffractive DIS region. The two models differ in the form of
the $Q^2$-dependence of the background contribution:
\begin{table}
   \begin{tabular}{|l|l|l|l|}
\hline
Parameter & Ref. \cite{bosted} & Model I & Model II \\
\hline
$c_1$  & 2.124 & 2.24 & 2.2 \\
\hline
$c_2$  & 2.569 & 2.73 & 2.73 \\
\hline
$c_4$  & 0.064 & 0.155 & 0.155 \\
\hline
$c_5$  & 0.549 & 0.549 & 0.7 \\
\hline
$c_6$  & 1.914 & 1.914 & 2.5 \\
\hline
$c_7$  & 1.0 & 1.0 & 1.5 \\
\hline
$A_T^7(0)$  & 3.419 & 5. & 5. \\
\hline
$A_L^7(0)$  & 11. & 15. & 15. \\
\hline
 \end{tabular}
\caption{The list of the resonance parameters and their values for
  Model I and Model II, as compared to the original fit of
  Ref. \cite{bosted} (see Table III of that reference). 
The notation of \cite{bosted} was kept.}
\label{tab0}
\end{table}
\begin{itemize}
\item
{\bf Model I}: 
The model used in \cite{ja_chuck} utilized the resonance parameters
obtained in \cite{bianchi} and the non-resonant Regge contribution
from \cite{regge_fit} that was fitted to the real photoabsorption data at
high energies. The $Q^2$-dependence of the high-energy part was taken
from the hybrid GVD/color dipole (CDP) approach of Ref. \cite{cvetic}. 
For the estimates of \cite{ja_chuck}, a simple dipole model with
the dipole mass $\Lambda\approx1$ GeV for all the transition resonance 
form factors was employed. Because it was found that this simple
dipole form fails dramatically throughout the resonance region, we
adopt the resonance part from \cite{bosted}
with a few parameters minimally adjusted in order to fit the data
with the background of a different form, rather the one used in
\cite{bosted} originally. We list those
parameters and the respective changes in Table \ref{tab0}.
\item
{\bf Model II}: 
To test the sensitivity of our calculations to the specific model, we use
another form of the background from the \lq\lq na\"ive'' GVD model of
Ref. \cite{alwall} (cf. Eqs. (3,4) of that Ref.), and we add the resonance contributions from \cite{bosted} on
top of that. Again, some resonance parameters are slightly adjusted to
the background, and all changes are quoted in Table \ref{tab0}. 
\end{itemize}
In Fig. \ref{fig:sigmatot} we confront the two models with the total 
photoabsorption cross section. 
The Model I is shown by solid red lines, Model II 
by the dashed blue line. 

\begin{figure}
\includegraphics[width=3in]{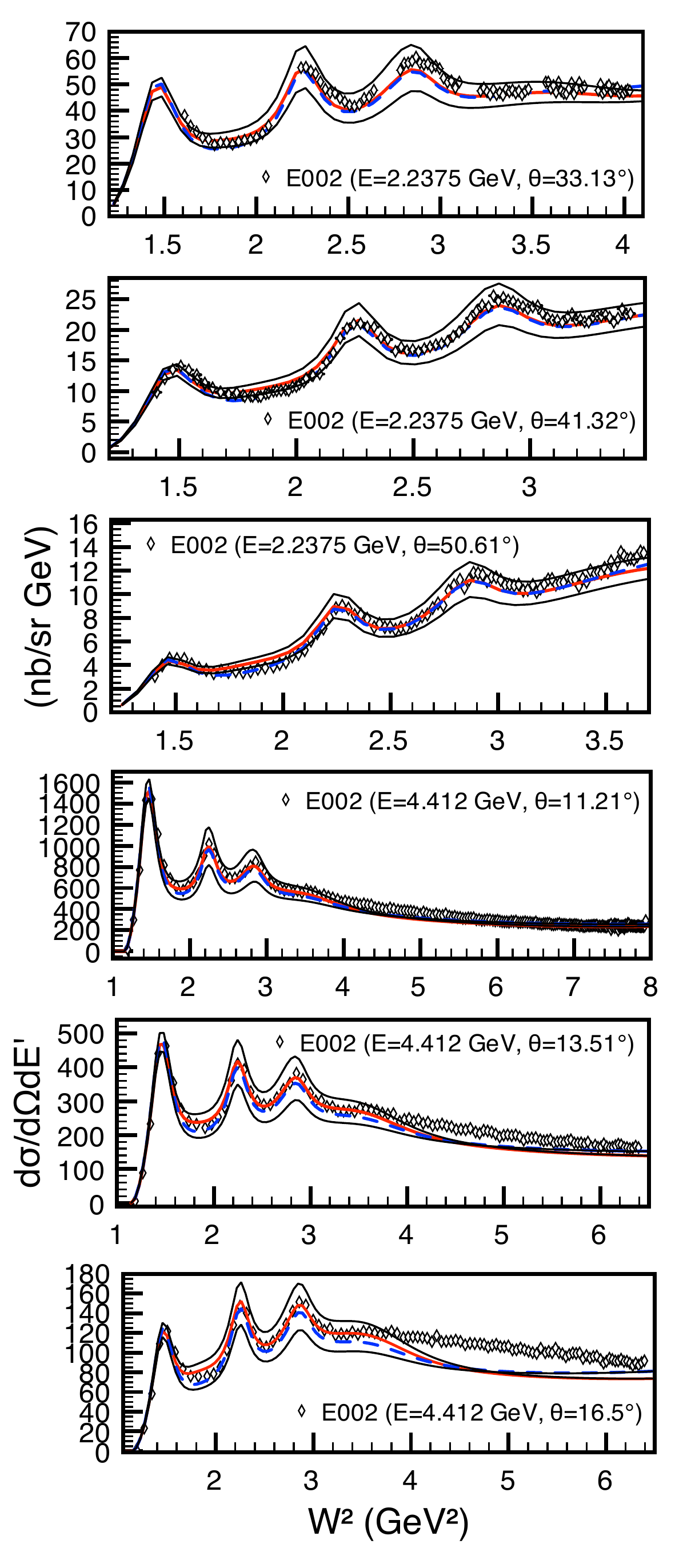}
\caption{(Color online) Differential cross section data in the resonance region from
  \cite{E00} are shown in comparison with the two models. The experimental errors are
  not shown. Thick solid line is the result of Model I, and thick
  dashed line is the result of Model II. Thin solid lines show the
  error bar due to the uncertanties in 
  helicity amplitudes for the photoexcitation of the resonances on the
  proton, according to \cite{PDG}.} 
\label{fig:resdata1}
\end{figure}

\begin{figure}
\includegraphics[width=3in]{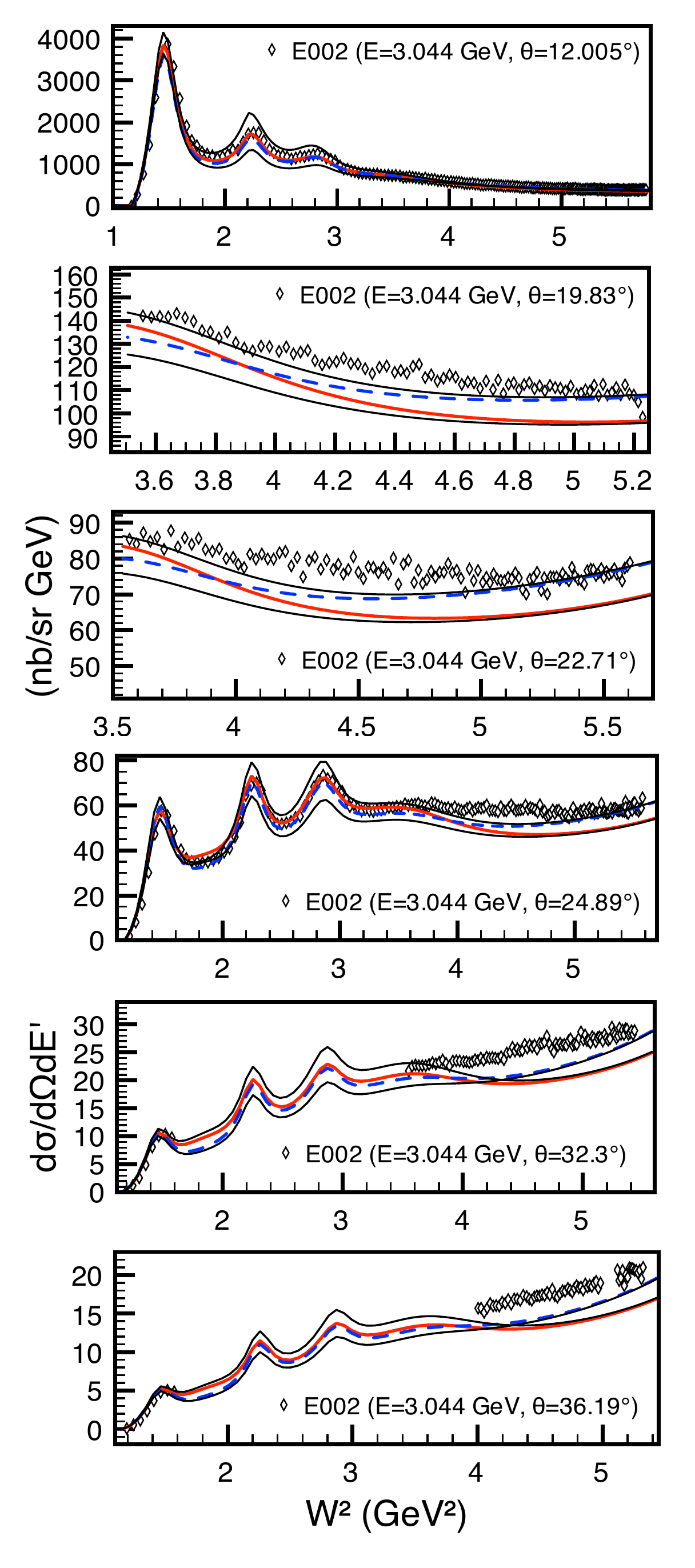}
\caption{(Color online) Differential cross section data in the resonance region from
  \cite{E00} are shown in comparison with Model I and II. Notation
  as in Fig. \ref{fig:resdata1}} 
\label{fig:resdata2}
\end{figure}

\begin{figure}
\includegraphics[width=3in]{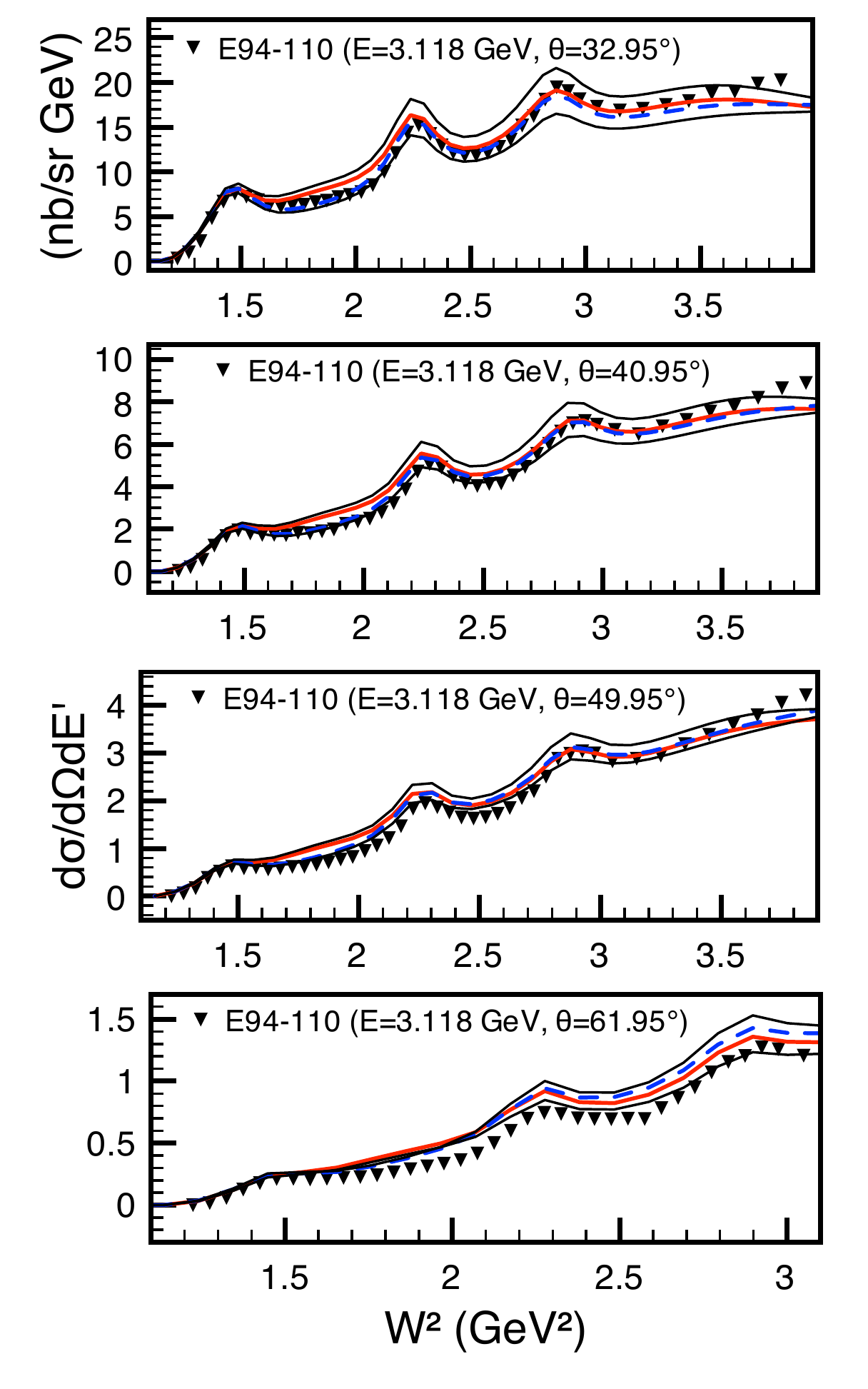}
\caption{(Color online) Differential cross section data in the resonance region from
  \cite{E94} are shown in comparison with Model I and II.  Notation
  as in Fig. \ref{fig:resdata1}} 
\label{fig:resdata6}
\end{figure}

Figs. \ref{fig:resdata1}-\ref{fig:resdata6} 
display the comparison
of the two models with the data for the differential cross section for
inclusive electroproduction in the resonance region. Both models in 
general provide a good description of the data in the resonance
region. The areas between the
lower and upper thin curves in each plot correspond to the range of
values of the helicity amplitudes for the photoexcitation of each resonance
included in Models I and II, as given by the PDG \cite{PDG}. It can be
seen that the experimental data are always contained within these
areas for $W^2\leq4$ GeV$^2$, even without including the experimental
errors. At the same time, we note that just above the resonance
region, in the limited range 4 GeV$^2\leq W^2\leq6$ GeV$^2$, and at
moderate values of $Q^2$, the background systematically lacks
strength. However, we stress that this lack of strength is observed
only in very limited range of energies, and the deficit is less than
20\% which makes the impact of this effect on the dispersion
correction small. 

\begin{figure}
\includegraphics[width=3in]{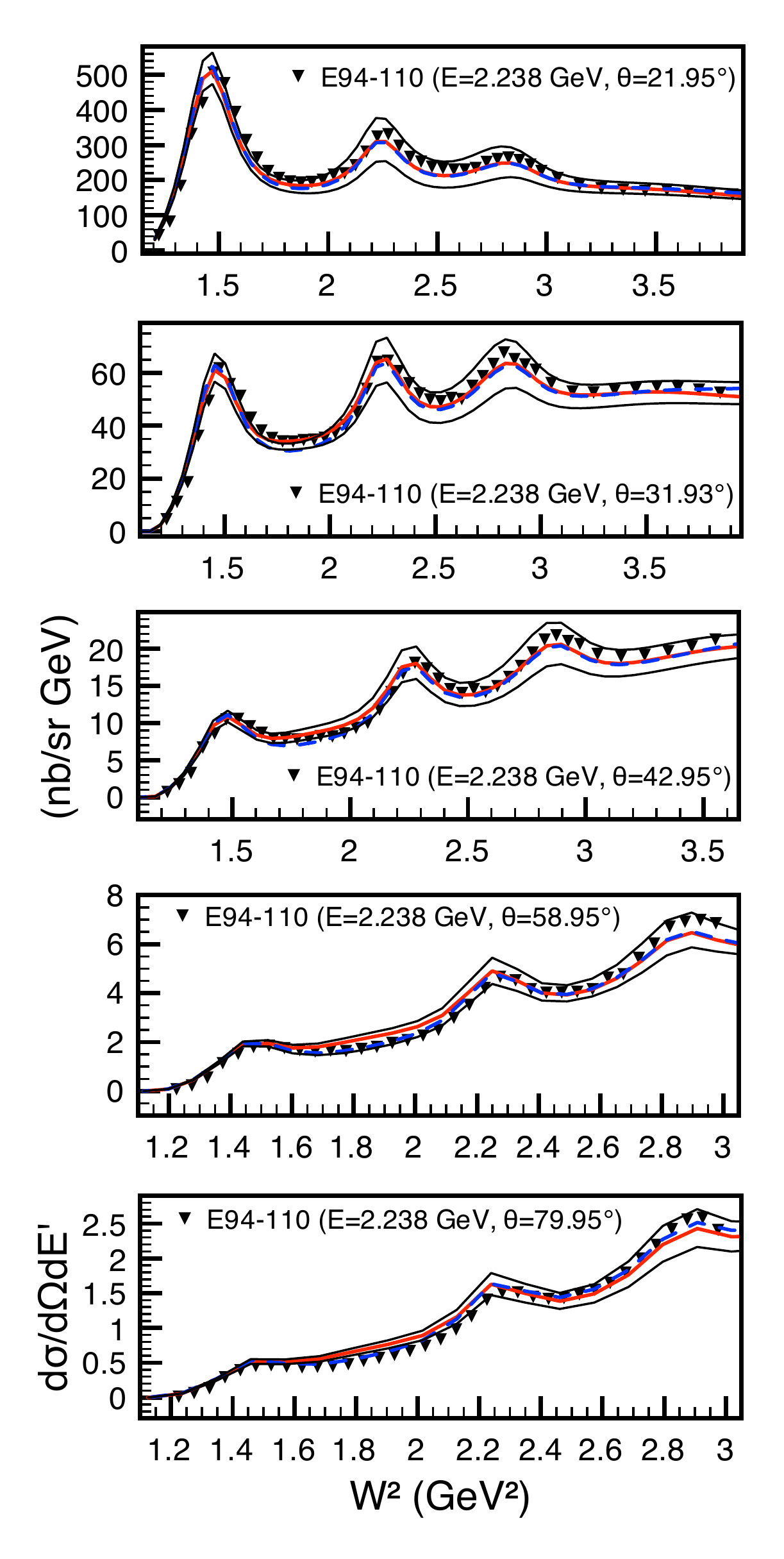}
\caption{(Color online) Differential cross section data in the resonance region from
  \cite{E94} are shown in comparison with Model I and II.  Notation
  as in Fig. \ref{fig:resdata1}} 
\label{fig:resdata3}
\end{figure}

\begin{figure}
\includegraphics[width=3in]{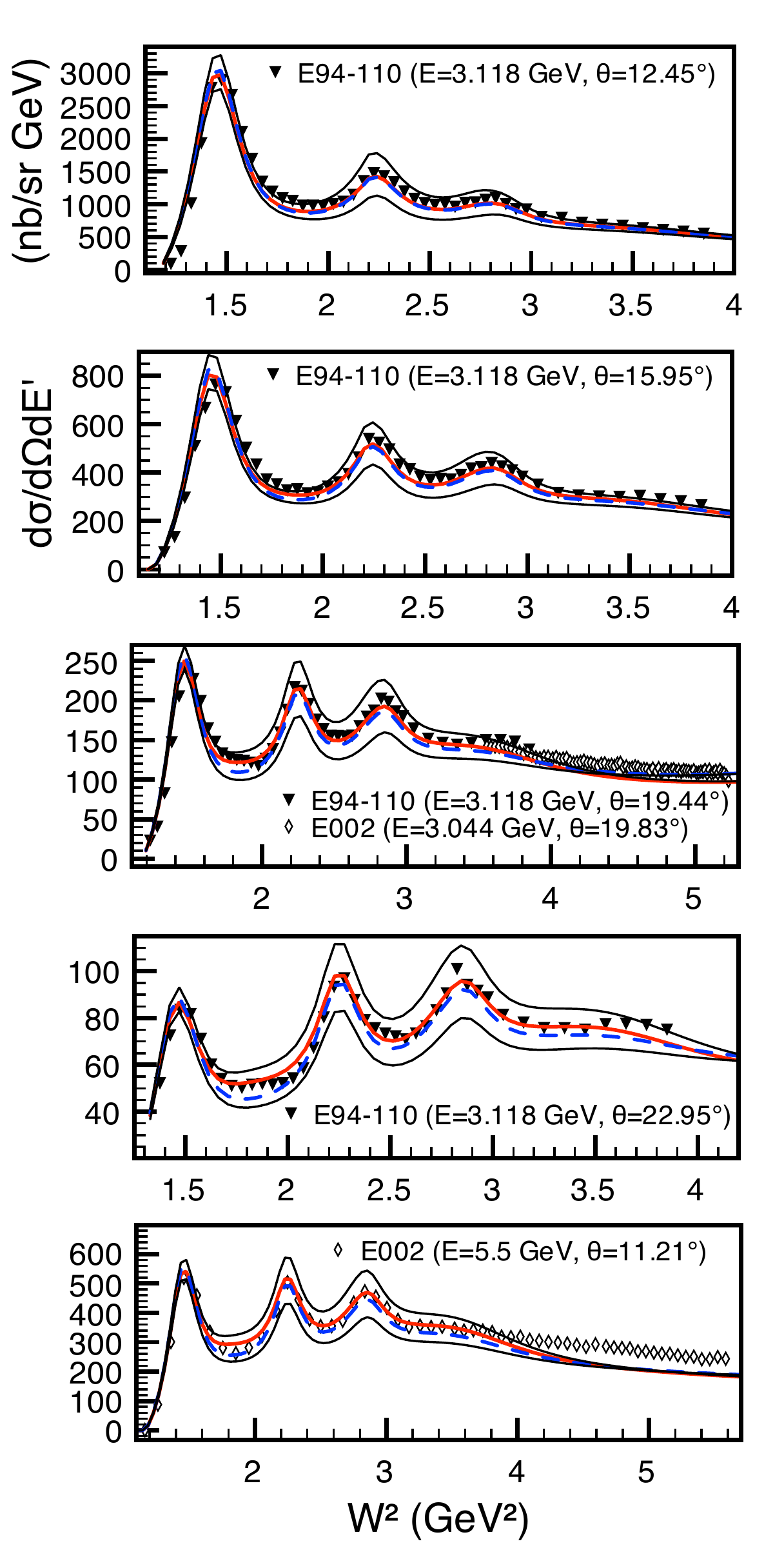}
\caption{(Color online) Differential cross section data in the resonance region from
  \cite{E94} are shown in comparison with Model I and II.  Notation
  as in Fig. \ref{fig:resdata1}} 
\label{fig:resdata4}
\end{figure}

\begin{figure}
\includegraphics[width=3in]{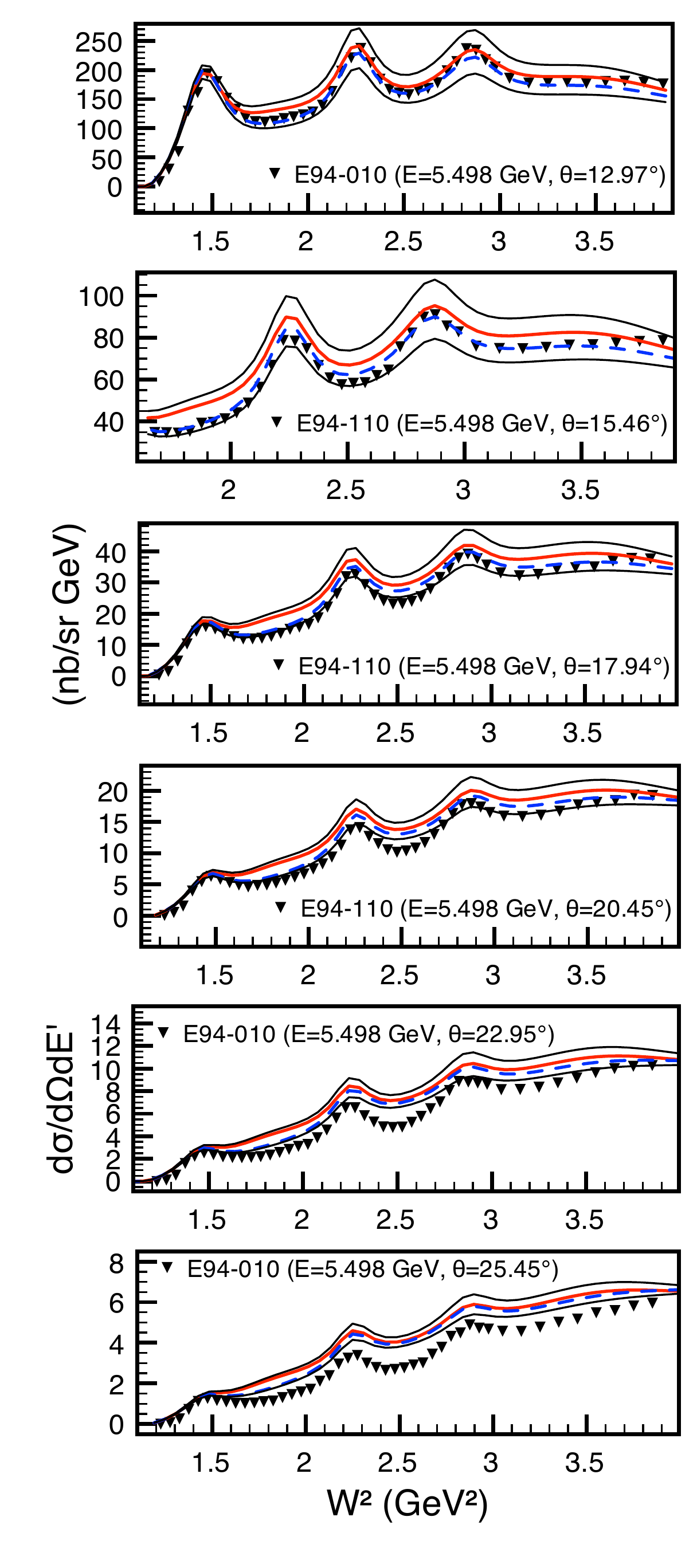}
\caption{(Color online) Differential cross section data in the resonance region from
  \cite{E94} are shown in comparison with Model I and II.  Notation
  as in Fig. \ref{fig:resdata1}} 
\label{fig:resdata5}
\end{figure}
\begin{figure}
\includegraphics[width=3in]{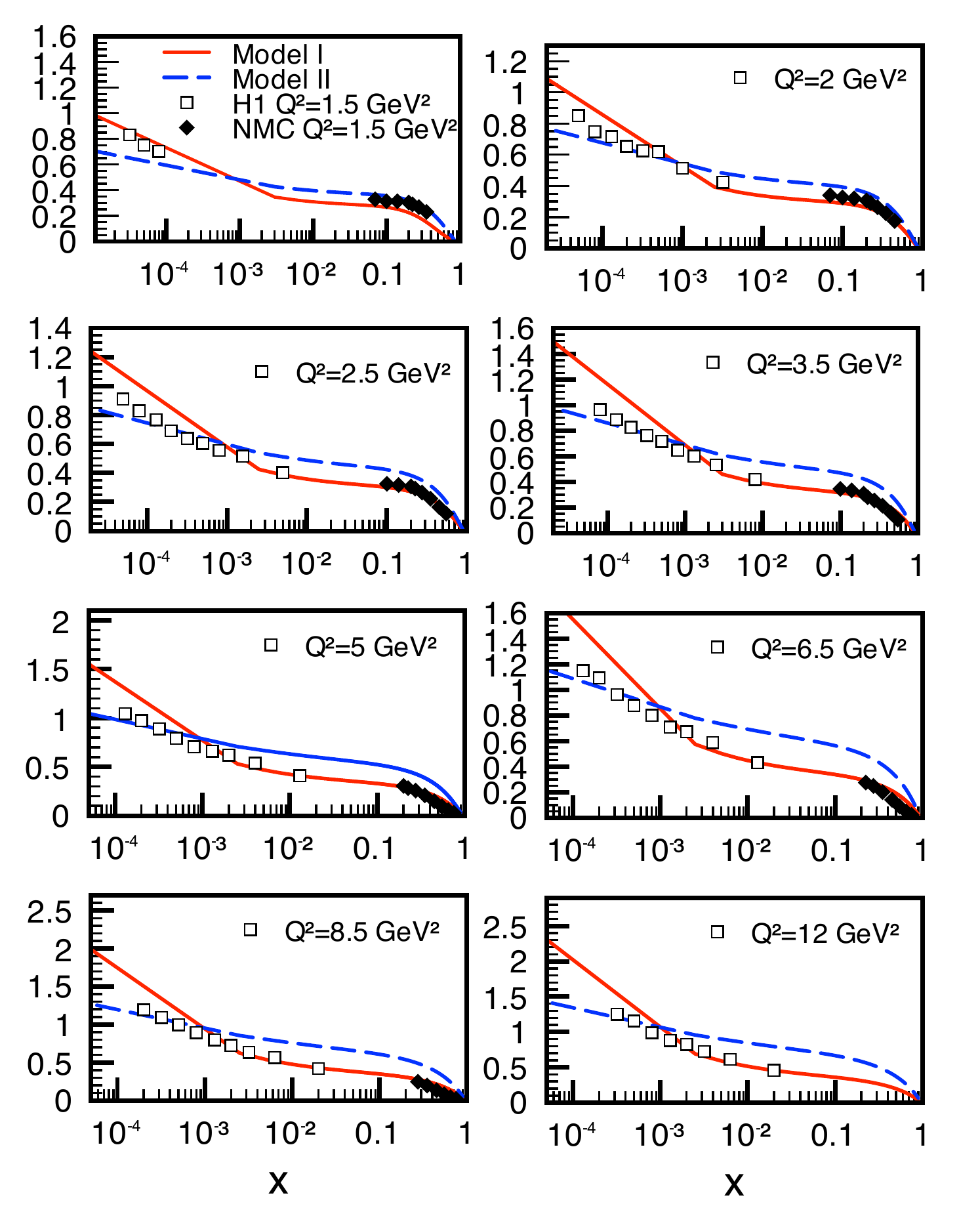}
\caption{Comparison of the $x$-dependence of the DIS structure function 
$F_2(x,Q^2)$ at fixed $Q^2$ and as a function of $x$, in
GVD/CDP model of \cite{cvetic} 
(solid lines) and the naive GVD model of \cite{alwall} (dashed lines) 
to the low-$x$ DIS data of H1 Collaboration \cite{F2_H1}.  The experimental errors are
  not shown.} 
\label{fig:F2vsx}
\end{figure}
\begin{figure}
\includegraphics[width=3in]{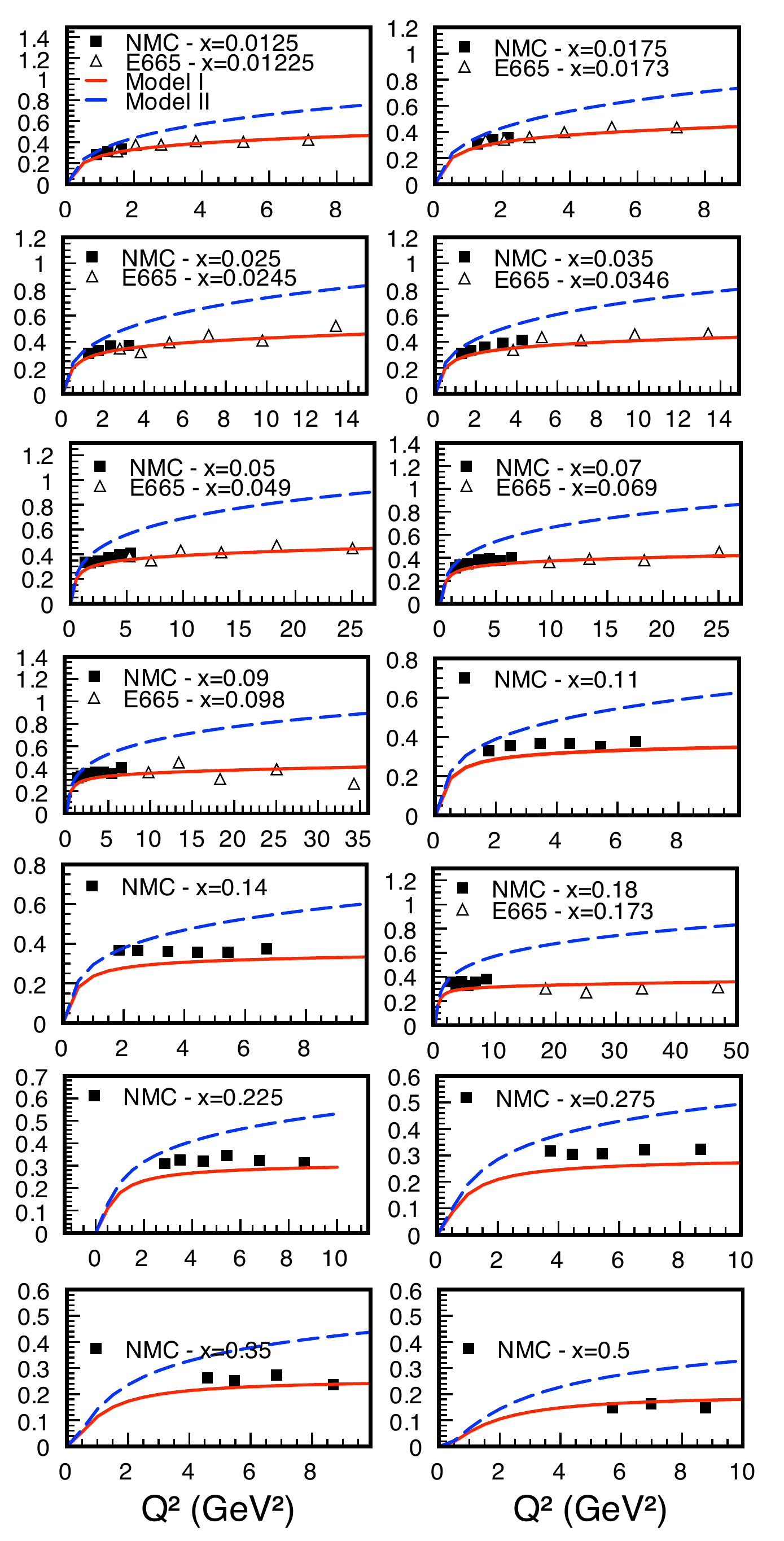}
\caption{Comparison of the $Q^2$-dependence of the DIS structure function 
$F_2(x,Q^2)$ at fixed $x$ and as a function of $Q^2$, in GVD/CDP model of 
\cite{cvetic} (solid lines) and the naive GVD model of \cite{alwall} 
(dashed lines) to the DIS data of NMC Collaboration \cite{F2_NMC} and 
E665 Collaboration \cite{E665} where the $x$-binning corresponds to that of 
NMC.  The experimental errors are
  not shown.}
\label{fig:F2vsQ2}
\end{figure}

We next turn to the deep inelastic (DIS) data. For DIS, a natural choice would
be to use the PDF parametrizations from MRST or CTEQ, DGLAP-evolved to
the necessary value of $Q^2$. However, this is 
only applicable at large enough $Q^2$, and
extrapolating them below $Q^2=1$ GeV$^2$ introduces additional
systematic error. 
In Figs. \ref{fig:F2vsx}, \ref{fig:F2vsQ2}, the na\"ive 
GVD model of Ref. \cite{alwall} (Model II) is shown 
along with the GVD/CDP model of \cite{cvetic} (Model I). 
One can see that while the GVD/CDP model reproduces the data in a wide range of 
$x,Q^2$, the naive GVD model overshoots the data at large $x$
starting at moderate $Q^2$, and underestimates the low-$x$ behavior for all 
$Q^2$. One needs to keep in mind, however, that both models work
reasonably well at moderate $Q^2$ and large $x$ which give the main
contributions to the dispersion correction.

The following comment is in order here. 
The authors of Ref. \cite{sibirtsev} argued that
our description of the data is unsatisfactory not only in the
resonance region but also beyond (cf. Fig. 1 of
\cite{sibirtsev}). While the model of the resonance form factors of
Ref. \cite{ja_chuck} was definitely not accurate (one of the instances
on which we improve that calculation in the present work), the model
for the background in \cite{ja_chuck} is exactly the same as that of
Model I here. We believe that Figs. \ref{fig:resdata1}-\ref{fig:F2vsQ2} presented in
this section provide abundant evidence of a satisfactory description of the
experimental data by our phenomenological model. In view of this, we
find it puzzling that Ref. \cite{sibirtsev} quotes a discrepancy of
40-50\% at $Q^2$ as low as 0.6 GeV$^2$ just above the resonance
region (cf. the upper left panel of Fig. 1 of that reference). 

\subsection{Isospin rotation of the resonance contributions}
In Standard Model, the $Z$ and $\gamma$ hadronic currents are related by 
means of a simple isospin rotation, 
\beqn
J_{em}^\mu&=&q^{I=0}J_{I=0}^\mu+q^{I=1}J_{I=1}^\mu+q^sJ_s^\mu\nn\\
J_{NC_V}^\mu&=&g_V^{I=0}J_{I=0}^\mu+g_V^{I=1}J_{I=1}^\mu+g_V^sJ_s^\mu,
\label{eq:iso}
\eeqn
with 
\beqn
J_{I=0}^\mu&=&\frac{1}{\sqrt{2}}(\bar u\gamma^\mu u+\bar d\gamma^\mu d)\nn\\
J_{I=1}^\mu&=&\frac{1}{\sqrt{2}}(\bar u\gamma^\mu u-\bar d\gamma^\mu d)\nn\\
J_{s}^\mu&=&\bar s\gamma^\mu s\ \ \ .
\eeqn
The e.m. charges given by 
\be
q^{I=0}=\frac{1}{3\sqrt{2}}, \quad
q^{I=1}=\frac{1}{\sqrt{2}},\quad
q^s=-\frac{1}{3}\ \ \ , 
\ee
whereas 
the weak charges are 
\bea
\nonumber
g_V^{I=0}&=&-\frac{1}{\sqrt{2}}\frac{4}{3}s^2\theta_W \\ 
\nonumber
g_V^{I=1}&=&\frac{1}{\sqrt{2}}(2-4s^2\theta_W) \\
\nonumber
g_V^s&=&-1+\frac{4}{3}s^2\theta_W\ \ \ ,
\eea
with $s^2\theta_W$ being a shorthand for $\sin^2\theta_W$ (for
purposes of this argument). 
This isospin decomposition is used to relate weak proton form factors
to the proton and neutron electromagnetic form factors,
\beqn
\langle p|J^\mu_{NC,V}|p\rangle=(1-4s^2\theta_W) \langle p|J^\mu_{em}|p\rangle
-\langle n|J^\mu_{em}|n\rangle 
\eeqn
where we neglected strangeness contributions that are generally small \cite{strangequarks}.

\bwt
\begin{table}
   \begin{tabular}{|l|l|l|l|l|l|l|l|}
\hline
&$P_{33}(1232)$ & $S_{11}(1535)$ & $D_{13}(1520)$ & $S_{11}(1665)$ &
$F_{15}(1680)$ & $P _{11}(1440)$& $F _{37}(1950)$\\
\hline
$y_R$ &-1.0$^{-0.1}_{+0.1}$ & -0.51$^{-0.71}_{+0.35}$ & -0.77$^{-0.125}_{+0.125}$ & -0.28$^{-0.86}_{+0.45}$ & -0.27$^{-0.12}_{+0.1}$ & -0.62$^{-0.2}_{+0.19}$ & -1$^{-1}_{+1}$\\
\hline
   \end{tabular}
\caption{Ratios $y_R$ with respective uncertainties for seven resonances.}
\label{tab}
\end{table}
\ewt

 The above relation is valid for transitions to $I=\frac{1}{2}$ resonances, as well: 
\beqn
\langle X|J^\mu_{NC,V}|p\rangle=(1-4s^2\theta_W) \langle X|J^\mu_{em}|p\rangle
-\langle X|J^\mu_{em}|n\rangle\,.
\eeqn
\indent
It is then straightforward to relate the contribution of a resonance
$R$ with isospin $1/2$ to the interference $\gamma Z$
cross section entering Eq.~(\ref{imdeltagz}) to its contribution to the electromagnetic cross section:
\beqn
\langle p|J^\mu_{em}|R\rangle\langle R|J^\mu_{NC,V}|p\rangle&=&
(1-4s^2\theta_W) |\langle R|J^\mu_{em}|p\rangle|^2\nn\\
&-&\langle p|J^\mu_{em}|R\rangle\langle R|J^\mu_{em}|n\rangle 
\label{eq:matrel_gz_gg_def}
\eeqn
\indent
Consequently, for each resonance, we define two ratios describing the relative
strength of its contribution to the $\gamma Z$-interference cross
sections $\sigma_{T(L),R}^{\gamma Z,p}$ with respect to the purely
electromagnetic ones  $\sigma_{T(L),R}^{\gamma p}$ as
\beqn
\xi_{Z/\gamma}^R(Q^2)&\equiv&\frac{\sigma_{T,R}^{\gamma Z,p}}
{\sigma_{T,R}^{\gamma p}}\nn\\
\zeta_{Z/\gamma}^R(Q^2)&\equiv&\frac{\sigma_{L,R}^{\gamma Z,p}}
{\sigma_{L,R}^{\gamma p}} 
\label{eq:xi_def}
\eeqn
\indent
In the Appendix A we discuss in detail the $Q^2$-dependence of these
ratios, as well as the ratios of the longitudinal cross sections $\zeta_{Z/\gamma}^R$. 
Basing on the discussion in Appendix A, we will use the value 
\beqn
\xi_{Z/\gamma}^R(Q^2)&=&\left[1-4s^2\theta_W(0)\right]-y_R=\mathrm{const.}\,,
\label{eq:xiZg}
\eeqn
to rescale the contribution of a resonance $R$ to both transverse and
longitudinal cross section. Possible discrepancies 
(which, if known, are model-dependent) from this rule are 
accounted for by assigning a conservative uncertainty to the ratios
$\xi_{Z/\gamma}^R$. This is done by using the PDG values and
respective errors for the transition helicity amplitudes. These PDG
values represent an average over
different data sets and different extraction procedures adopted in the
various experiments. Consequently, they automatically 
include an enhanced error due to model dependence of this extraction. 

The first term in Eq. (\ref{eq:xiZg}) is a constant that is model-independent, arising from Eq. (\ref{eq:matrel_gz_gg_def}). This model independence reflects the cancelation of the proton-to-resonance transition matrix elements involving the e.m. currents.
The second term in Eq. (\ref{eq:xiZg}) , 
$y_R$, is given by the ratio of combinations of neutron and 
proton transverse helicity amplitudes (we refer the reader to the Appendix
for  details).
We summarize the values of $y_R$ obtained using the PDG values for
 the helicity amplitudes with the respective errors in Table
 \ref{tab}. The lower and upper limits correspond to taking extreme values of the
transition helicity amplitudes for the proton and neutron from
\cite{PDG}. 

For the $P_{33}(1232)$ resonance, we assign a conservative 10\% error
on its isospin structure. According to the PDG, this error should be
precisely zero. However, the analyzes of Refs. \cite{bosted,bosted2}
return slightly different results for the $P_{33}(1232)$ excitation on
the proton and the neutron, both for real and virtual photons. The 
discrepancy stays below relative 10\% for $Q^2\leq1$ GeV$^2$, although 
this conclusion is definitely model-dependent. This observation provides 
motivation for assigning a conservative 10\% error to $y_R$ for the $P_{33}(1232)$.

Similarly, for the $F _{37}(1950)$ resonance, the uncertainty is
driven by the analyses of Refs. \cite{bosted,bosted2}. The fit of
\cite{bosted} for the proton returns a very mild monopole form factor,
whereas the neutron data require a dipole form factor for the same
resonance \cite{bosted2}. Also the strength strongly depends on the
form of the background, as found in our work (see Table
\ref{tab0}). This motivated us to assign a conservative 100\%
uncertainty due to this resonance. 

We note that for both $S_{11}$ resonances
listed in the Table \ref{tab} the error bar exceeds 100\%. This is
mostly due to the quality of the extracted values for the neutron. 
It is also worth noting that the quark model expectations (see
Table I of Ref. \cite{deltagZ_conf} for the isospin scaling factors
within the quark model of Ref. \cite{koniuk}) are not too far from
the central values quoted in Table \ref{tab}. 

\subsubsection{Uncertainty in isospin rotating the
 resonances}
To summarize the results of the previous subsection, 
we propose to obtain the contribution of a resonance $R$ to the $\gamma
Z$-interference cross sections $\sigma_{T,L}^{\gamma Z, R}$ by multiplying
the purely electromagnetic cross sections
$\sigma_{T,L}^{\gamma\gamma,R}$ with a scaling factor
$\xi^R_{Z/\gamma}$ that is {\it independent of $W^2$ and
  $Q^2$}. Furthermore, to the precision required here, we rescale the
trasverse and longitudinal cross sections with the same factor. 
Each such factor contains two parts, as per Eq.~(\ref{eq:xiZg}):
the first one is model-independent, whereas the second one is obtained
from the analysis of the proton and neutron electromagnetic data, and
involves model dependence and experimental uncertainties. 
The values of $y_R$ are listed in Table \ref{tab} with the respective
uncertainties. 
Correspondingly, for each resonance we simply obtain its contribution
to the interference structure functions $F_{1,2}^{\gamma Z,R}$ from that
to the electromagnetic structure functions $F_{1,2}^{\gamma\gamma,R}$ as 
\beqn
F_{1,2}^{\gamma Z,R}(W^2,Q^2)&=&\xi^R_{Z/\gamma}F_{1,2}^{\gamma\gamma,R}(W^2,Q^2)
\label{F12_rescale}
\eeqn

To compute Re$\,\Box_{\gamma Z_A}$, we use Eqs. (\ref{eq:dr}) and
(\ref{imdeltagz}) with the input from Eqs. (\ref{F12_rescale}) and
(\ref{eq:f12-sigmaTL}). Finally, we use the parametrizations of the 
transverse and longitudinal electromagnetic cross sections from 
Model I and Model II, and values of $\xi^R_{Z/\gamma}$ factors from
Table \ref{tab}. The uncertainty on the contribution of each resonance
is obtained according to the definition 
\beqn
\Delta F_{1,2}^{\gamma Z,R}(W^2,Q^2)&=&\Delta y_R F_{1,2}^{\gamma\gamma,R} (W^2,Q^2),
\label{F12R_uncertainty}
\eeqn
where $\Delta y_R$ are the uncertainties quoted in Table \ref{tab}.
Using the steps described above for the individual contributions of
resonances to Re$\,\Box^R_{\gamma Z_A}$, we can also compute the
uncertainties $\Delta\,($Re$\,\Box^R_{\gamma Z_A})$ associated with each such contribution. 
Because most resonances do not overlap, we treat all these 
uncertainties as independent, thus we define 
\beqn
\Delta_R^{Full}{\rm Re}\,\Box_{\gamma Z}&=&\sqrt{\sum_R |\Delta\,({\rm Re}\,\Box^R_{\gamma Z_A})|^2}
\label{F12R_uncertainty_full}
\eeqn
\subsection{Isospin rotation of the high energy contribution}
We need to employ a well-motivated model to describe the 
isospin dependence of the background contribution. One option is to employ the
the VDM picture, incorporating the simple observation that the 
photon has the same quantum numbers as vector mesons (VM). Therefore, it can 
fluctuate into $\rho,\omega$ or $\phi$ that then scatter off the
nucleon. This approach underlies the background in both Models I and II, so we proceed generally at first.

According to the VDM, the photon can be represented as a superposition of a few vector mesons,
\beqn
|\gamma\rangle=\sum_{V=\rho,\omega,\phi}\frac{e}{f_V}
|V\rangle,
\eeqn
with $f_V$ the VM decay constant. 
Assuming this basis to be complete and orthogonal (no VM mixing), 
one can express the total photoabsorption cross section
through a combination of total cross sections for vector meson-proton
scattering, 
\beqn
\sigma_{tot}(\gamma p)&=&\sum_V\frac{4\pi\alpha}{f_V^2}\sigma_{Vp}
\eeqn
At high energies, the total cross section $\sigma_{Vp}$ should be
independent of the VM flavor and the above equation becomes simply a
flavor decomposition of the electromagnetic total cross section, 
although this representation is of limited use because $\sigma_{Vp}$
is unknown. Nevertheless, 
after trivial manipulations this picture leads to the VDM 
(Stodolsky) sum rule \cite{stodolsky} that relates the total, real photoabsorption 
cross section to a sum of differential cross sections for photoproduction of vector 
mesons, 
\beqn
\sigma_{tot}(\gamma p)&=&\sum_{V=\rho,\omega,\phi}
\sqrt{16\pi\frac{4\pi\alpha}{f_V^2}\frac{d\sigma^{\gamma p\to Vp}}{dt} (t=0)}.
\label{eq:vdmsumrule}
\eeqn
\indent
This sum rule is based on the assumptions of vector meson dominance and 
almost purely imaginary phase of scattering amplitudes at high energy. 

In the na\"ive GVD approach (Model II), this sum rule holds only approximately (HERA data: $\approx$80\%) 
\cite{stodolsky_hera}. The missing strength can be attributed to the neglect of non-diagonal vector meson-nucleon 
scattering $Vp\to V'p$. One can then generalize the VDM by including such 
contributions by writing down a dispersion relation over the vector meson 
masses. We will denote this non-diagonal \lq\lq continuum" contribution as \lq\lq $X$" in the sum over
vector mesons $V$.  

Alternately, in the GVD/CDP approach (Model I), instead of hadronic VM states, the photon hadronic
wave function (WF) is described in terms of perturbative $q\bar q$-states
with $J=1$. This $q\bar q$-pair forms a color dipole that interacts 
with the target through gluon exchanges. 

Both the na\"ive GVD and GVD/CDP approaches are similar in the following
instances: they consider the interaction of the hadron-like photon
with the target (hadronic WF for na\"ive GVD, and perturbative $q\bar q$ for
GVD/CDP), and the interaction of the hadronic states is independent of
flavor (either VM or quark). 
This allows us to cast the ratio of inclusive
virtual photon and $\gamma Z$-interference cross sections in the following form:
\beqn
\sigma_{\gamma^*p}&=&
\sum_{V=\rho^0,\omega,\phi,X}r_{\gamma^*V}(W^2,Q^2)\sigma_{Vp},\nn\\
\sigma_{\gamma^*Zp}&=&\sum_{V=\rho^0,\omega,\phi,X}r_{\gamma^*ZV}(W^2,Q^2)\sigma_{Vp} \ \ \ ,
\eeqn
where \lq\lq $X$" denotes the non-diagonal contribution.

\indent
According to the assumptions of both approaches, 
the flavor factors $r_{\gamma^*V}(W^2,Q^2)$ and $r_{\gamma^*ZV}(W^2,Q^2)$
only contain the information about the projectile (virtual photon or
$Z$) and not about the target; this means that they cannot depend on
the energy but only on $Q^2$ (the only Lorentz scalar that can be
constructed from the $\gamma$ four-momentum) and the flavor of the
VM state. 
However, if these flavor factors indeed depend on energy, this would signal the
breakdown of the models, and would be a source of an additional theory
uncertainty. For completeness, we will keep the $W^2$ dependence. 
The interference flavor factors $r_{\gamma^*ZV}$ obtain 
from the purely electromagnetic ones using the 
conservation of the vector current (CVC), 
\bwt
\beqn
r_{\gamma^*Z\rho}(W^2,Q^2)&=&\frac{g_V^{I=1}}{e_q^{I=1}}r_{\gamma^*\rho}(W^2,Q^2)
=(2-4\sin^2\theta_W)r_{\gamma^*\rho}(W^2,Q^2),\nn\\
r_{\gamma^*Z\omega}(W^2,Q^2)&=&\frac{g_V^{I=0}}{e_q^{I=0}}r_{\gamma^*\omega}(W^2,Q^2)
=-4\sin^2\theta_Wr_{\gamma^*\omega}(W^2,Q^2),\nn\\
r_{\gamma^*Z\phi}(W^2,Q^2)&=&\frac{g_V^{s}}{e_q^{s}}r_{\gamma^*\phi}(W^2,Q^2)
=(3-4\sin^2\theta_W)r_{\gamma^*\phi}(W^2,Q^2),
\eeqn
for the light flavors. 

With these definitions, 
we obtain our master formula for rescaling the background
contribution:
\beqn
\frac{\sigma_{T,L}^{\gamma^*p\to Zp}}{\sigma_{T,L}^{\gamma^*p\to\gamma^*p}}
&=&
\frac{(2-4\sin^2\theta_W) r^{T,L}_{\gamma^*Z\rho}(W^2,Q^2)-4\sin^2\theta_Wr^{T,L}_{\omega}(W^2,Q^2)
+(3-4\sin^2\theta_W)r^{T,L}_{\phi}(W^2,Q^2)+r^{T,L}_{\gamma^*ZX}(W^2,Q^2)}
{r^{T,L}_{\gamma^*\rho}(W^2,Q^2)
+r^{T,L}_{\gamma^*\omega}(W^2,Q^2)
+r^{T,L}_{\gamma^*\phi}(W^2,Q^2) 
+r^{T,L}_{\gamma^*X}(W^2,Q^2)}\nn\\
&=&
\frac{(2-4\sin^2\theta_W)-4\sin^2\theta_WR^{T,L}_{\frac{\omega}{\rho}}(W^2,Q^2)
+(3-4\sin^2\theta_W)R^{T,L}_{\frac{\phi}{\rho}}(W^2,Q^2)+\frac{r^{T,L}_{\gamma^*ZX}}{r^{T,L}_{\gamma^*\rho}}}
{1+R^{T,L}_{\frac{\omega}{\rho}}(W^2,Q^2)+R^{T,L}_{\frac{\phi}{\rho}}(W^2,Q^2)+R^{T,L}_{\frac{X}{\rho}}(W^2,Q^2)}.
\label{eq:master}
\eeqn
\ewt
\indent
The ratios $R^{T,L}_{\frac{V}{\rho}}$ are defined as ratios of
transverse ($T$)  or longintudinal ($L$) vector meson ($V$) production cross sections
\be
R^{T,L}_{\frac{V}{\rho}}=\frac{\sigma^{\gamma^*p\to Vp}_{T,L}}{\sigma^{\gamma^*p\to\rho p}_{T,L}}\ \ \ .
\ee
The terms $\sim r^{T,L}_{\gamma^*X},\,r^{T,L}_{\gamma^*ZX}$ account for the
possible incompleteness of the VDM (or three light flavor) basis. 

For the na\"ive GVD model, $R^{T,L}_{\frac{V}{\rho}}$ are obtained from the
experimentally measured constants $f_V$ of the leptonic decay $V\to
e^+e^-$. Additionally, the presence of the VM propagator leads to a
prediction for the $Q^2$-dependence of each flavor channel 
$\sim\left(\frac{m_V^2}{m_V^2+Q^2}\right)^2$, thus we have 
\beqn
R^T_{\frac{V}{\rho}}=R^L_{\frac{V}{\rho}}=\frac{\sigma^{\gamma^*p\to Vp}}{\sigma^{\gamma^*p\to\rho p}}=
\frac{f_\rho^2}{f_V^2}\frac{m_V^4}{m_\rho^4}
\left(\frac{m_\rho^2+Q^2}{M_V^2+Q^2}\right)^2
\label{eq:vdm_mesonprod}
\eeqn
with $V=\omega,\phi$. The remaining piece, $R^{T,L}_{\frac{X}{\rho}}$ is
identified with the continuum ($V-V'$ mixing) contribution. From the comparison
of the left and right hand sides of the VDM sum rule
\cite{stodolsky_hera} and suplementing this contribution with a simple $Q^2$
dependence to describe the virtual photoabsorption data at low and
moderate $Q^2$, one obtains i.e. for the transverse ratio \cite{alwall}
\beqn
R^T_{\frac{C}{\rho}}&=&R_{\frac{C}{\rho}}(0)\frac{(1+Q^2/m_\rho^2)^2}{(1+Q^2/m_0^2)}
\label{eq:vdm_cont}
\eeqn
with $R_{\frac{C}{\rho}}(0)=\frac{0.21}{0.67}$ and $m_0\approx1.5$
GeV. We note that due to the monopole $Q^2$-dependence of
the continuum contribution, rather than dipole for the $\rho^0$, the
impact of the continuum part increases with growing $Q^2$.
The master formula of Eq. (\ref{eq:master}) -- together with the model
input of Eqs. (\ref{eq:vdm_mesonprod}) and (\ref{eq:vdm_cont}) (see
Ref. \cite{alwall} for all the details of the model)  -- 
defines our prescription for the isospin rotation of the background
contribution within the na\"ive GVD model (Model II).

For pQCD inspired models, such as the GVD/CDP used in Model I, the
relative strength of the isospin (flavor) channels is directly related
to the quark electric charges and is independent of energy and
$Q^2$. 
\bwt
\beqn
\sigma^{\gamma\to\rho}\;:\;
\sigma^{\gamma\to\omega}\;:\;
\sigma^{\gamma\to\phi}\;:\;
\sigma^{\gamma\to J/\psi}&=&
1\;:\;\frac{(q^{I=0})^2}{(q^{I=1})^2}\;:\;
\frac{(q^{s})^2}{(q^{I=1})^2}\;:\;
\frac{(q^{c})^2}{(q^{I=1})^2}=1\;:\;\frac{1}{9}\;:\;\frac{2}{9}\;:\;\frac{8}{9}
\label{eq:su4_mesonprod}
\eeqn
\ewt
One possible way is to identify the $X$ state in the master
formula with the $c\bar c$ state, i.e. $J/\psi$. In that case, the $X$
contribution in the numerator of Eq. (\ref{eq:master}) is given
according to the SM
\be
\frac{r^{T,L}_{\gamma^*ZX}}{r^{T,L}_{\gamma^*\rho}}=\frac{3-8\sin^2\theta_W}{2} 
R^{T,L}_{\frac{J/\psi}{\rho}}\ \ \ .
\ee
The choice of identifying $X$ with $J/\psi$ is justified in HERA
kinematics but is probably less convincing at lower energies and 
low $Q^2$. Moreover, the choice $X=J/\psi$ and the relative
strength of different contributions according to
Eq. (\ref{eq:su4_mesonprod}), corresponds to
the VDM sum rule being saturated to only 60\%, rather than the
measured 80\%, suggesting that it is not very realistic. 

Either way, for the rescaling of the background contribution in the GVD/CDP model (Model I), 
Eqs. (\ref{eq:master}) and (\ref{eq:su4_mesonprod}) simply combine to
a constant factor. Its value when using only the three light flavors
amounts to
\beqn
\label{eq:modInocharm}
\left[\frac{\sigma^{\gamma^*p\to
      Zp}}{\sigma^{\gamma^*p\to\gamma^*p}}\right]^\mathrm{Model\;I}_{u,d,s}
=2-4\sin^2\theta_W\approx1.05 
\eeqn
\indent
However, when including the charm contirbution, one obtains
\beqn
\label{eq:modIcharm}
\left[\frac{\sigma^{\gamma^*p\to Zp}}{\sigma^{\gamma^*p\to\gamma^*p}}\right]^\mathrm{Model\;I}_{u,d,s,c}
=\frac{9}{5}-4\sin^2\theta_W\approx0.85
\eeqn
\indent
For comparison, a typical value of this ratio within the na\"ive GVD
Model II (we quote its value at $Q^2=0$ for definiteness: in Model II
it is $Q^2$-dependent, although mildly) is
\beqn
\left[\frac{\sigma^{\gamma^*p\to
      Zp}}{\sigma^{\gamma^*p\to\gamma^*p}}\right]^\mathrm{Model\;II}_{\rho,\omega,\phi}
\approx1.92-4\sin^2\theta_W\approx0.97 \,,
\eeqn
and a very similar number when including the continuum and assuming
its size for the $\gamma Z$ cross section to be equal to that for the
purely electromagnetic case. 
However, any such estimate bears at least 20\% uncertainty due to the
incompleteness of the na\"ive VDM basis and due to the unknown flavor
structure of the continuum contribution.

\begin{figure}
\includegraphics[width=8.6cm]{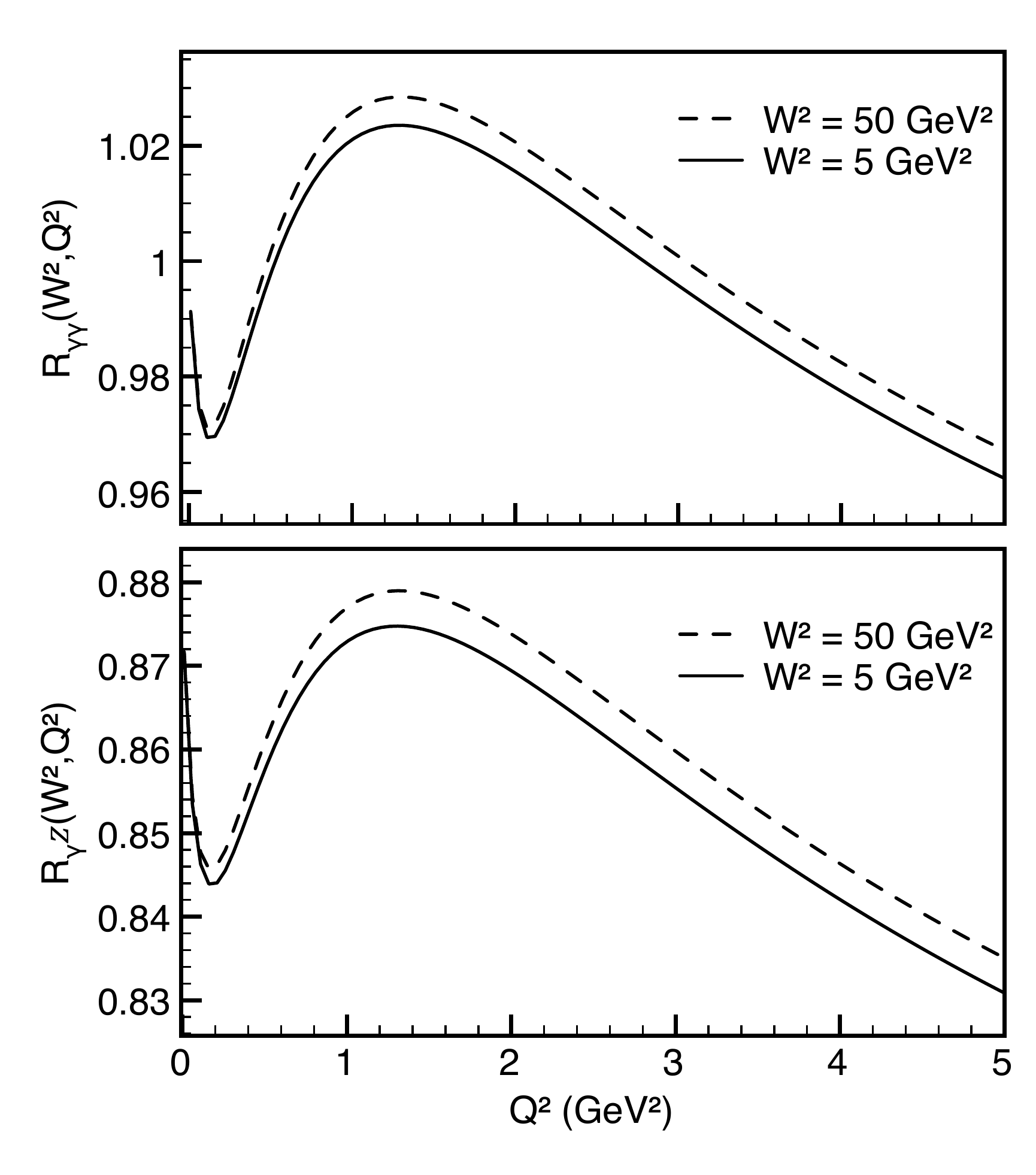}
\caption{Ratios $R_{\gamma\gamma}(W^2,Q^2)$ (upper panel) and
  $R_{\gamma Z}(W^2,Q^2)$ (lower panel) are shown as function of $Q^2$
at $W^2=5$ GeV$^2$ (solid lines) and at $W^2=50$ GeV$^2$ (dashed
lines). See text for further details.} 
\label{fig:VDM_GVD_ratio}
\end{figure}

To illustrate the difference in the $Q^2$-dependence of the total
cross section 
as calculated in Model I and Model II, we define the
following two ratios,
\beqn
R_{\gamma\gamma}(W^2,Q^2)&=&\frac{\Big[\sigma_T^{\gamma^*p\to\gamma^*p}+\sigma_L^{\gamma^*p\to\gamma^*p}\Big]^{\rm Model\,I}}{\Big[\sigma_T^{\gamma^*p\to\gamma^*p}+\sigma_L^{\gamma^*p\to\gamma^*p}\Big]^{\rm Model\,II}}\nn\\
R_{\gamma Z}(W^2,Q^2)&=&\frac{\Big[\sigma_T^{\gamma^*p\to
    Zp}+\sigma_L^{\gamma^*p\to Zp}\Big]^{\rm
    Model\,I}}{\Big[\sigma_T^{\gamma^*p\to
    Zp}+\sigma_L^{\gamma^*p\to Zp}\Big]^{\rm Model\,II}}\,,\nn\\
\eeqn
where we suppressed the arguments of the cross sections for
compactness. 
In Fig. \ref{fig:VDM_GVD_ratio} we display the $Q^2$-dependence of
$R_{\gamma\gamma}$ and $R_{\gamma Z}$ at two values of $W^2$. The
ratios show very mild $W^2$-dependence, in accord with general
assumptions used in VDM and GVD/CDP models. The $Q^2$-dependence shows
slight oscillations (at the level of 3\%) at $Q^2\lesssim2$ GeV$^2$;
at higher values of $Q^2$ both ratios decrease monotonically, as a
result of the na\"ive VDM model (Model II) overshooting high-$Q^2$
data significantly, while GVD/CDP Model I describes data in a wide
kinematical range.

\subsubsection{Uncertainty in isospin rotating the background}
\begin{figure}
\includegraphics[width=8.6cm]{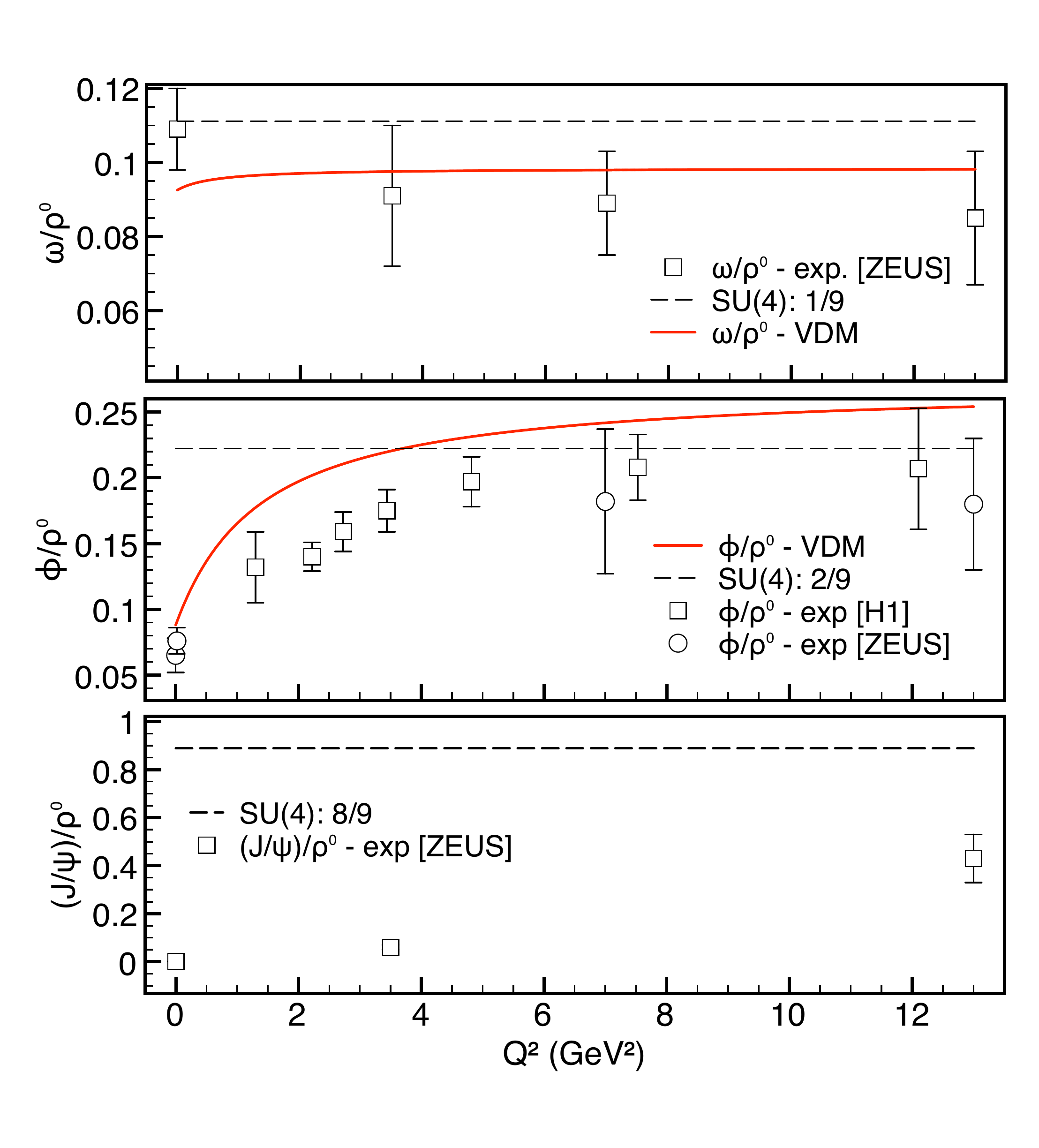}
\caption{(Color online) Experimental data for ratios of total cross sections for
  elastic vector meson electroproduction in comparison with the naive
  VDM (solid lines) and perturbative SU(4) (dashed lines) predictions.} 
\label{fig:V-rho_ratios}
\end{figure}

We wrote our master formula in terms of ratios of meson production
cross sections, rather than cross sections themselves. These ratios
were recently measured at HERA.
The predictions of Eqs. (\ref{eq:vdm_mesonprod},\ref{eq:su4_mesonprod})
are confronted with the experimental data of Ref. \cite{rho_omega_phi} at high
energies and for $Q^2$ that ranged from zero to several GeV$^2$ in Fig. \ref{fig:V-rho_ratios}.
To estimate uncertainties in isospin rotation of Models I and II, we will
directly compare the model predictions of the isospin ratios to the
HERA data. The common feature of the two models is that these ratios are
$W^2$-independent. 
Furthermore, Model I predicts them to be $Q^2$-independent, too.
Instead, Model II (na\"ive GVD) predicts the $Q^2$-running of these ratios.
In both cases, we will 
assume that the uncertainty in isospin scaling the $I=1$ channel
(i.e., the $\rho^0$) is zero. 
For each flavor channel, we define the uncertainty as the discrepancy 
\beqn
\label{eq:errordef}
\Delta\frac{\sigma^{\gamma\to V}}{\sigma^{\gamma\to \rho}}(Q^2)=
\left(\frac{\sigma^{\gamma\to V}}{\sigma^{\gamma\to \rho}}\right)^\mathrm{Model}
-\left(\frac{\sigma^{\gamma\to V}}{\sigma^{\gamma\to
      \rho}}\right)^\mathrm{exp}\!\!\!, 
\eeqn
with $V=\rho,\omega$ for VDM and $V=\rho,\omega,(J/\psi)$ for pQCD. 
Additionally, for VDM we assign a 100\% uncertainty to the continuum
contribution whose flavor content is not defined in the naive GVD approach.
Similarly, for GVD/CDP model we assign a conservative 100\% uncertainty
to the $c\bar c$ contribution, in view of an unsatisfactory description
of the data for $(J/\psi)/\rho$ by the $SU(4)$ prediction. 

Arriving at the estimate of the total uncertainty due to the isospin
structure of the background requires following steps. 
For Models I and II, we insert the uncertainties defined in Eq.~(\ref{eq:errordef}) and below for each flavor $V$ into the master formula
Eq. (\ref{eq:master}).  Subsequently, we use the relation of
Eq. (\ref{eq:f12-sigmaTL}) and obtain the total uncertainty of the interference structure
functions, $\Delta F_{1,2}^{\gamma Z}$. We evaluate the imaginary part of the dispersion
  correction of Eq. (\ref{imdeltagz}) with 
$\Delta  F_{1,2}^{\gamma Z}$. The final step involves evaluating 
the dispersion integral thereof, Eq. (\ref{eq:dr}). 
These steps give us the uncertainties due to the isospin structure of
the background within Model I and Model II. 
To be conservative, we will choose the larger of the two as our estimate of the non-resonant
model uncertainty. 

Anticipating the discussion in the next Section, we note that the overall uncertainty 
is dominated by the continuum contribution (\lq\lq  $X$")  within the naive GVD model. 
The only significant assumption about the continuum contribution here
is that its size (relative to diagonal vector meson contributions) is
energy-independent , and we take it from the data at very high
$W^2$. Until now, the only dedicated study of the VDM sum rule was
performed at $W\geq70$ GeV at HERA -- far from the kinematic region that dominates the dispersion integral
for $\Box_{\gamma Z}$. It is not {\em a priori} clear that the decomposition of the virtual photon
into the VM basis works any differently for 5 GeV photons than for 80
GeV photons. 

As part of a program of future measurements to constrain the uncertainties in the dispersion 
correction, it would be useful to have direct data on this sum rule at lower energies: $2\leq W\leq10$ GeV. 
In case that new data on the VDM sum rule at these 
energies will become available, it will then be straightforward to include
additional $W$-dependent form factors in Eq. (\ref{eq:master}). Data on the virual vector meson photoproduction cross sections in this kinematic regime could also provide additional important constraints. Together with direct measurements fo the inelastic PV asymmetries at these kinematics, such measurements could in principle lead to a significant reduction in the quoted theoretical error bar.

\section{Results for Re$\Box_{\gamma Z}$}
\begin{figure}
\vspace{0.5cm}
\includegraphics[width=8.6cm]{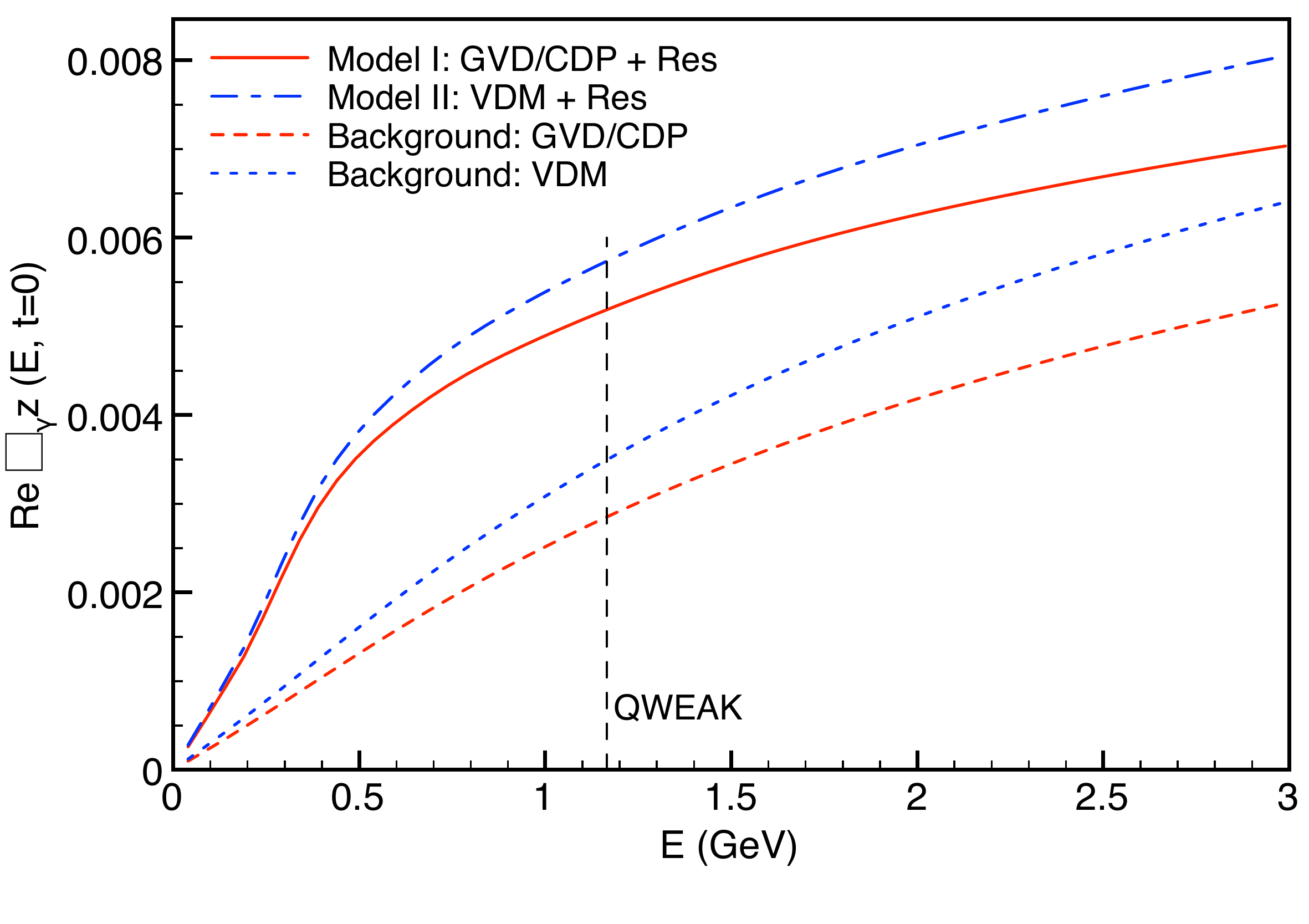}
\caption{(Color online) Sum of resonance and background contributions to Re $\Box_{\gamma Z}$
  for Models I and II as indicated in the legend of the plot.} 
\label{fig:deltagZ_full}
\end{figure}

We are now in the position to present results for $\Box_{\gamma Z}$
in the forward direction using the sum rule of
Eqs. (\ref{imdeltagz},\ref{eq:dr}), the Models I and II for the
electromagnetic cross sections along with the isospin considerations
provided in the previous sections. 
We display the sum of resonance and background in
Fig. \ref{fig:deltagZ_full}. 

\begin{figure}
\includegraphics[width=8.6cm]{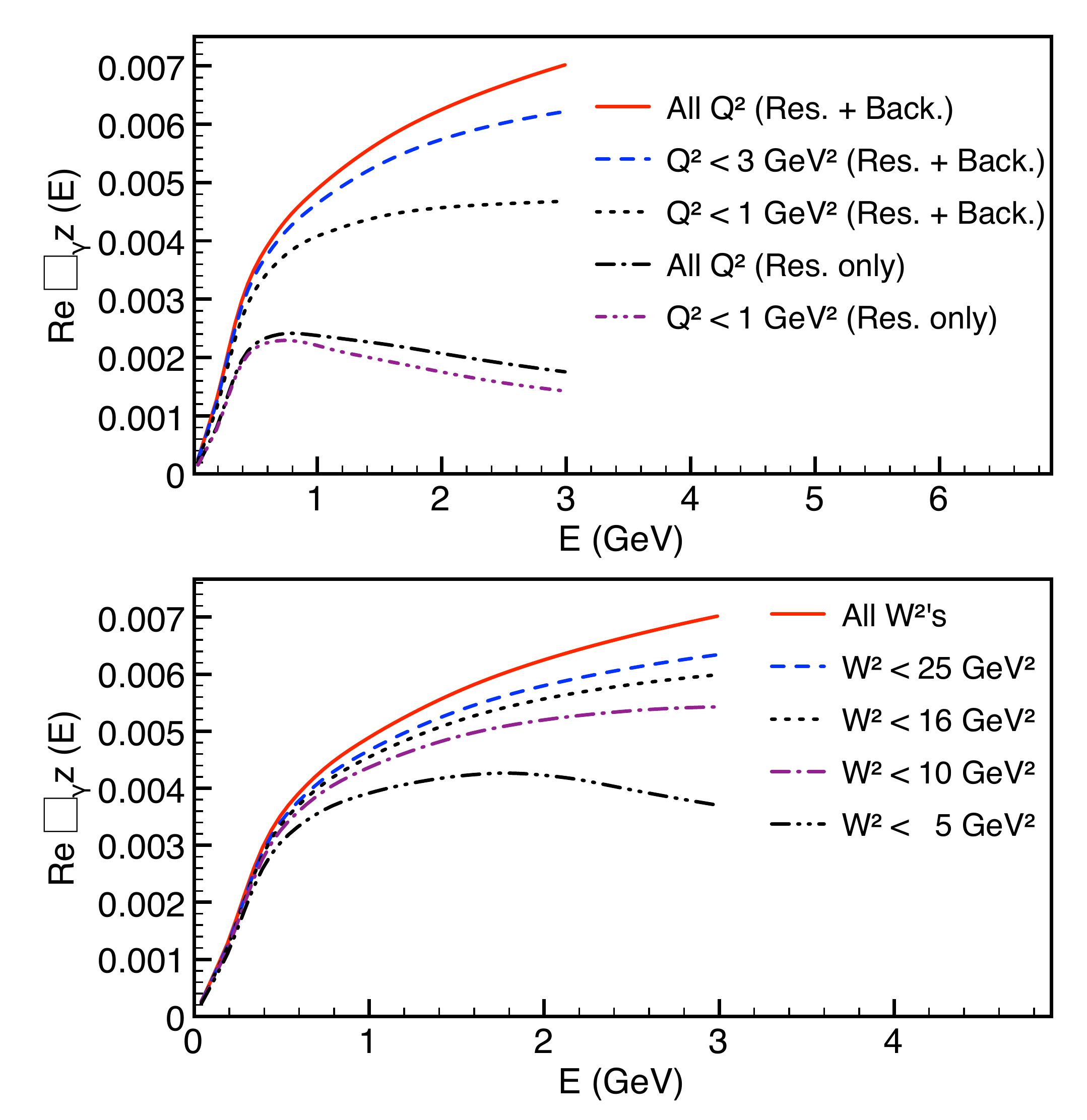}
\caption{(Color online) Contributions of different kinematic regions to $\Box_{\gamma Z}$
  for Models I. In the upper panel, contributions from various $Q^2$
  ranges are shown: solid curve (full result), dashed line
  ($Q^2<3$ GeV$^2$ for resonance + background), dotted line
  ($Q^2<1$ GeV$^2$ for resonance + background). For comparison, in the
  upper panel we display result of integration of the resonance
  contribution to $\Box_{\gamma Z}$ over all values of $Q^2$
  (dash-dotted curve) and $Q^2<1$ GeV$^2$ (dash-double-dotted curve).
In the lower panel, contributions to $\Box_{\gamma Z}$ are shown that
come from $W^2\leq5$ GeV$^2$ (dash-double-dotted curve), 
$W^2\leq10$ GeV$^2$ (dash-dotted curve), $W^2\leq16$ GeV$^2$ (dotted
curve), $W^2\leq25$ GeV$^2$ (dashed curve), and full result (solid curve).} 
\label{fig:kinreg}
\end{figure}

In Fig. \ref{fig:kinreg}, we display the contributions of various
kinematic regions to Re $\Box_{\gamma Z}$. 
The upper panel of
Fig. \ref{fig:kinreg} evidences that the resonance contribution is
dominated by values of $Q^2\leq1$ GeV$^2$, whereas for the total
correction, values of $Q^2$ up to 3 GeV$^2$ have to be taken to
saturate the dispersion correction to $\approx90$\%. The lower panel
of that Figure demonstrates that values of $W^2$ up to 25 GeV$^2$ have
to be included under the integration to saturate the dispersion
correction Re $\Box_{\gamma Z}$. The data from the resonance region
$W^2\leq5$ GeV$^2$ (resonance plus background) only contribute about
65\% of the total. The notation $Q^2\leq Q_A^2$ and $W^2\leq W_B^2$
refers to evaluating the double integral for Im $\Box_{\gamma
  Z}$ in Eq. (\ref{imdeltagz}) only over those values of $Q^2\,(W^2)$ that lie
below $Q_A^2\,(W_B^2)$, respectively. After that, the dispersion
integral of Eq. (\ref{eq:dr}) is evaluated without further modifications. 

In Table \ref{tab4}, we display the background contribution as
calculated in Model I and II for the QWEAK kinematics. It can be seen
that the background represents both the largest contribution and the
source of the largest uncertainty. Most notably, within the naive GVD approach (Model II),
it is completely dominated by the continuum contribution whose isospin
structure is undetermined. In the pQCD approach (Model I), a
contribution similar in strength is assigned to the $c\bar c$ state. However,
because in this case we know exactly how the weak boson couples to
$c$-quarks, the uncertainty is about half the size of that for Model
II. This 50\% reduction is simply due to the fact that 
$g_V^c\approx\frac{1}{3}=\frac{1}{2}e_c$. 
\begin{table}
   \begin{tabular}{|l|l|}
\hline
& Background \\
\hline
Model I  &($2.85\pm0.85$) $\times10^{-3}$\\
\hline
Model II  &($3.49\pm1.92$)$\times10^{-3}$ \\
\hline
   \end{tabular}
\caption{Background contribution to the
  dispersion correction to the weak charge of the proton
  Re$\Box_{\gamma Z}$ at the QWEAK energy $E_{lab}=1.165$
  GeV. Results for Model I and Model II are shown.}
\label{tab4}
\end{table}

The individual resonance contributions are displayed in Table \ref{tab3}.
It can be seen that the overall
uncertainty in the resonance contribution is dominated by the
uncertainty in two contributions, namely $S_{11}(1535)$ and 
$F_{37}(1950)$. The former, in turn is dominated by the uncertainty in
the neutron transition helicity amplitude. The heavy resonance state is
not well-determined and should be studied in greater detail to
decrease the respective uncertainty for the dispersion correction. 
\bwt
\begin{table}
   \begin{tabular}{|l|l|l|l|l|l|l|l|l|}
\hline
&$P_{33}(1232)$ & $S_{11}(1535)$ & $D_{13}(1520)$ & $S_{11}(1665)$ & $F_{15}(1680)$ &
$P _{11}(1440)$& $F _{37}(1950)$ & $\sum$ Res.\\
\hline
Model I ($\times10^{-3}$)&($1.21\pm0.12$) & ($0.28^{+0.34}_{-0.17}$) &
($0.18\pm0.03$) & ($0.06^{+0.14}_{-0.06}$)
& ($0.04^{+0.013}_{-0.011}$) & ($0.09\pm0.03$) & ($0.48\pm0.44$)  &  ($2.34^{+0.59}_{-0.50}$)\\
\hline
Model II  ($\times10^{-3}$) &($1.23\pm0.12$) & ($0.29^{+0.34}_{-0.17}$) &
($0.18\pm0.03$) & ($0.06^{+0.14}_{-0.06}$) 
& ($0.04^{+0.013}_{-0.011}$) & ($0.06\pm0.02$) & ($0.40\pm0.36$)  &  ($2.24^{+0.53}_{-0.43}$)\\
\hline
   \end{tabular}
\caption{Resonances contributions to the
  dispersion correction to the weak charge of the proton
  Re$\Box_{\gamma Z}$ at the QWEAK energy $E_{lab}=1.165$
  GeV, in units of $10^{-3}$. For each contribution, we indicate the uncertainty discussed in
  the text. Results for Model I and Model II are shown.}
\label{tab3}
\end{table}
\ewt

\begin{figure}
\includegraphics[width=8.6cm]{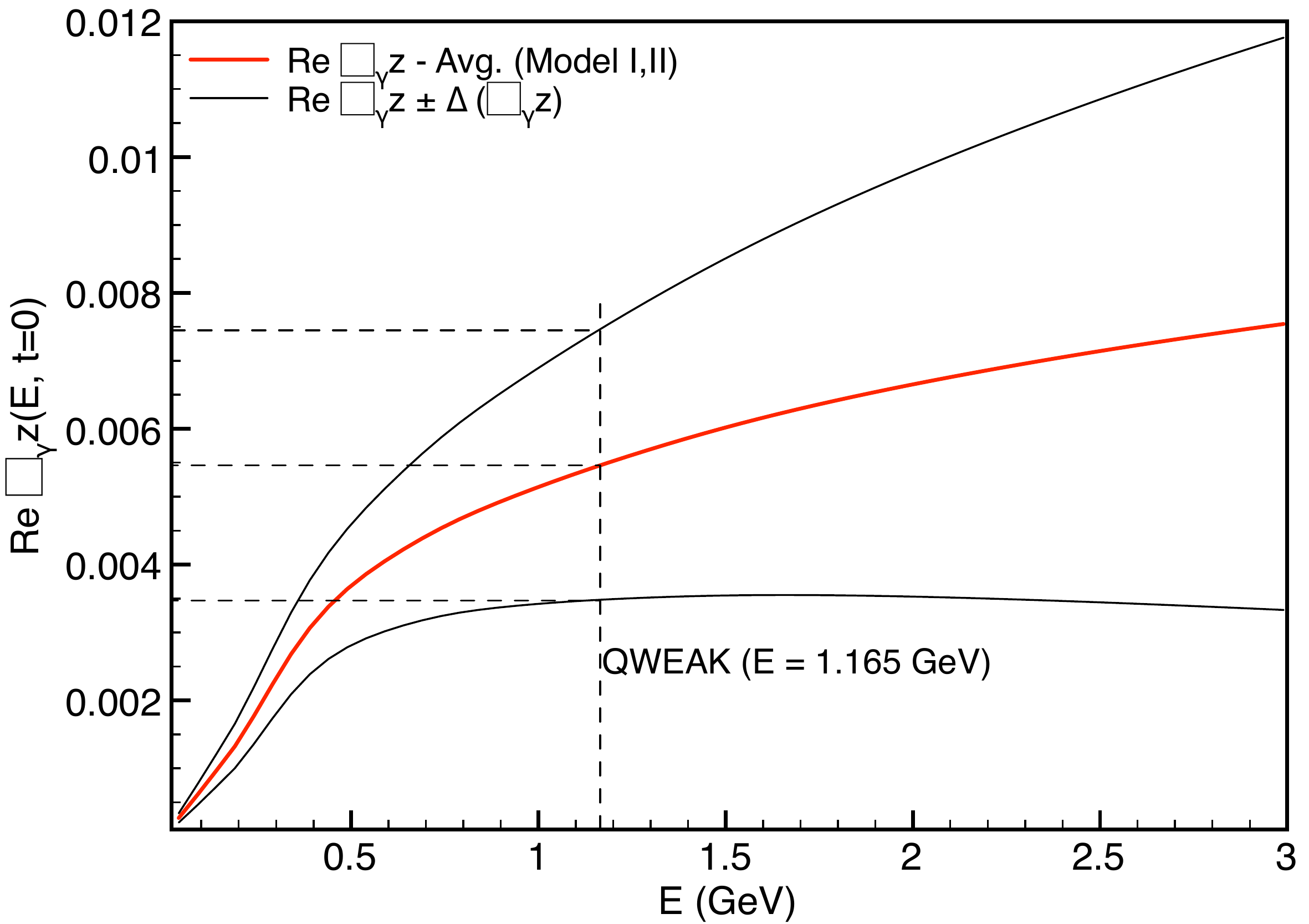}
\caption{(Color online) Full result for Re $\Box_{\gamma Z}$ with the theoretical
  error bar.} 
\label{fig:deltagZ_error}
\end{figure}

According to the discussion in the previous Section, we plot the
result for Re $\Box_{\gamma Z}$ and display the error bar on this
calculation in Fig. \ref{fig:deltagZ_error}. For the central value, we
take the average
of Model I and Model II, and use the difference between this central
value and either of Model I or II as the uncertainty due to modeling
the e.-m. data. For the isospin rotation-related uncertainty, we calculate the error
within each model as discussed before, and quote the larger of the
two. 
We summarize this section by quoting the result of the forward sum
rule evaluated within two models as follows:
\bwt
\beqn
{\rm Re}\,\Box_{\gamma Z}(E=1.165\,{\rm
  GeV},t=0)&=&[5.46\pm0.27\,(\mathrm{mod.\,avg.})\,\pm1.92\,(\mathrm{backgr.})^{+0.59}_{-0.50}\,(\mathrm{res.})]\times10^{-3},
\label{estimate}
\eeqn
The first uncertainty is due to averaging over the two models, the
second due to uncertainty in isospin rotating the background, and the
third due to isospin rotation of resonances. A possibility of
measuring 
the proton's weak charge at Mainz at a lower energy $E_{lab}=180$ MeV
is under consideration presently 
\cite{maas}, and we quote our prediction for the dispersion $\gamma Z$ 
correction and the respective uncertainty for that energy,
\beqn
{\rm Re}\,\Box_{\gamma Z}(E=0.180\,{\rm
  GeV},t=0)&=&[1.32\pm0.05\,(\mathrm{mod.\,avg.})\,\pm0.27\,(\mathrm{backgr.})^{+0.11}_{-0.08}\,(\mathrm{res.})]\times10^{-3}\,.
\label{eq:deltagZ_mami}
\eeqn
\ewt
We see that the total uncertainty in ${\rm Re}\,\Box_{\gamma Z}$ is
about six times smaller at $E_{lab}=180$ MeV than at $E_{lab}=1.165$
GeV. 

\section{Additional $t$-dependence of dispersion corrections}
In the previous section, we provided an educated estimate for Re
$\Box_{\gamma Z}$ in the exact forward direction. However, real
experiments are carried out at finite momentum transfer $t$, in
particular $|t|=0.03$ GeV$^2$ for the kinematics of the QWEAK
experiment. To extrapolate the forward sum rule to non-zero momentum
transfer, we employ the phenomenological model that was successfully
used for the beam normal spin (Mott) asymmetry in elastic $ep$-scattering 
\cite{afanasev,javvcs,ja_chuck_lead}. This model is inspired by i)
experimental data on the Compton differential cross section at small $t$
and high energy, and ii) the assumption of the predominantly imaginary
phase of the Compton amplitude at high energies (as for the pomeron). 

The data exhibit an exponential $t$-dependence,
\beqn
\frac{d\sigma}{dt}&=&\left(\frac{d\sigma}{dt}\right)_{t=0}\;e^{-B|t|}
\eeqn
with the slope parameter $B=7\pm1$ GeV$^{-2}$ \cite{aleksanian}. 
The differential cross section is related to the Compton amplitude
squared, whereas the total cross section -- through the optical theorem -- is related  
to the imaginary part of the Compton amplitude. Naively, then, one might expect the
$t$-dependence of the total cross section near the forward scattering limit to be close to
half as rapid as that of the differential cross section. Based on this {\em ansatz}, 
Ref. \cite{afanasev} proposed parameterizing
the $t$-dependence of the slightly off-forward total cross section
as 
\beqn
\sigma_\mathrm{tot}(t)&\approx&\sigma_\mathrm{tot}(t=0)\;e^{-\frac{B|t|}{2}}\ \ \ .
\eeqn
This parameterization becomes precise at very high energies where the cross section is
pomeron-dominated. We will follow a similar parameterization here.

This intrinsic $t$-dependence of the $\gamma Z$-box
contribution should be combined with the $\gamma\gamma$-box
contribution that becomes non-zero when going to finite
$t$. 
We found the effect of the dispersive contributions to $\Box_{\gamma\gamma}$ on $A^{PV}$ to be
negligibly small, of order below $0.1$\% at the QWEAK kinematics at
$-t=0.03$ GeV$^2$. The reason for this smallness is 
due to an explicit $t$-suppression of $\Box_{\gamma\gamma}$ with
respect to the tree level PC amplitude. 
Using the same approach, we obtain for the $t$-dependence of the dispersion correction  
\beqn
\Box_{\gamma Z}(E, t)&=&\Box_{\gamma
  Z}(E,0)\frac{exp(-B|t|/2)}{F_1^{\gamma p}(t)},
\eeqn
according to the definition of $\Box_{\gamma Z}$ as the ratio of the 
$\gamma Z$-box contribution to the PV amplitude $f_4(E,t)$ to the 
elastic proton electromagnetic form factor $F_1^{\gamma p}(t)$.

\begin{figure}
\vspace{0.5cm}
\includegraphics[width=8.6cm]{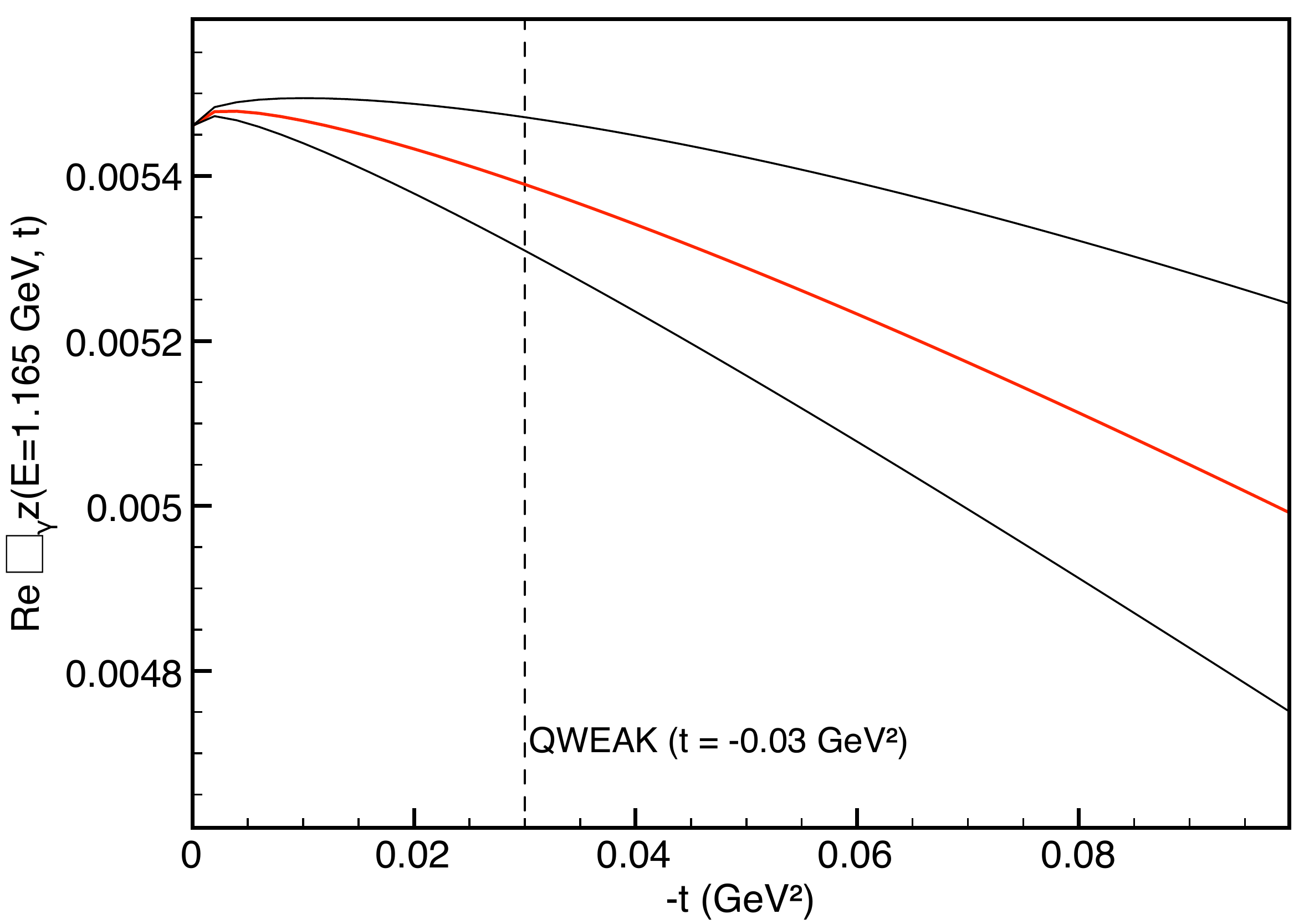}
\caption{(Color online) The dependence of the combined dispersion correction to
 $Q_W^p$ on the elastic momentum transfer $t$. The central value is
 shown in red; the thin black lines indicate the uncertainty.} 
\label{fig:tdep}
\end{figure}
In Fig. \ref{fig:tdep}, we display the $t$-dependence of the
combined dispersion correction for small values of the elastic
momentum transfer. It can be seen that one can expect that at
$|t|=0.03$ GeV$^2$, the dispersion correction decreases by only 
about 2\% relative to its value at $t=0$, and the same is valid for
the uncertainty in calculating this correction. We emphasize, however, that the model
for the $t$-dependence is derived from high-energy Compton
data, and it is not necessarily applicable to the resonance contributions. Thus, our estimate of
the effect of the $t$-dependence should be considered as an exploratory investigation. 
Having this caveat in mind and taking into account this $t$-dependence, we
obtain our final result for the dispersion $\gamma Z$ correction at
the kinematics of the QWEAK experiment:
\bwt
\bea
\label{eq:final}
{\rm Re}\,\Box_{\gamma Z_A}(E=1.165\,\mathrm{GeV},t =
  -0.03\, \mathrm{GeV}^2)&=&\\
\nonumber
\Bigl[ 5.39\pm0.27\, (\mathrm{mod.\, avg.})&\pm&1.88\,(\mathrm{backgr.})^{+0.58}_{-0.49}\,(\mathrm{res.})\pm0.07\,(t-\mathrm{dep.}) \Bigr]\,\times\, 10^{-3}
\ .
\eea
\ewt

To assess the relative impact of the energy-dependent contribution from $\Box_{\gamma Z_A}$ we first quote the result from Refs.~\cite{erler,erler2} for the weak charge, as defined in Eq.~(\ref{eq:qweakdef}):
\beqn
Q_W^p =0.0713\pm 0.0008
\eeqn
\indent
Compared to this prediction, 
the relative effect of the  $\Box_{\gamma Z_A}$ contribution at the
kinematics of the QWEAK experiment is
\beqn
\frac{{\rm Re}\,\Box_{\gamma Z_A}}{Q_W^p}
=(7.6\pm2.8)\%
\eeqn
Because this contribution was initially neglected in the analysis of
radiative corrections for the QWEAK experiment, the final theory
prediction and the respective uncertainty have to be corrected to include it.
Treating all the individual uncertainties quoted above as
independent, we obtain
\bea
\lim_{t\to 0} \frac{W^\mathrm{PV}}{W^\mathrm{EM}}&=&\qwp+\Box_{\gamma Z_A}\nn\\
&=&(0.0767\pm0.0008\pm0.0020_{\gamma Z})\,.
\eeqn
\indent
As discussed in Refs.~\cite{erler,erler2}, the first error includes a
conservative $\sim 1\%$ 
error associated with
$C_{\gamma Z}$ that appears in the prediction for $\qwp$. 
The additional uncertainty associated with the energy-dependent
contribution 
from $\Box_{\gamma Z_A}$ is almost
entirely due to the uncertainty due to the isospin structure of the
background. The latter, in turn, is largely dominated (70\%) by the
uncertainty due to the ``continuum'' contribution that does not have
well-defined isospin content. 

To recollect, the continuum contribution
that arises in  both the na\"ive GVD and GVD/CDP frameworks is a measure of the incompleteness of the
vector meson basis for the energetic photon. Its value at the
real photon point is obtained from the direct comparison of the VDM sum rule with the experimental photoproduction cross sections
where one finds roughly a 20\% deficit in the na\"ive GVD approach. Departing from the real photon
point, one employs phenomenological models for the $Q^2$-dependence
for this contribution, such that in this way the generalized VDM
description fits the virtual photoabsorption data at moderate $Q^2$. 
It turns out that above $Q^2=2$ GeV$^2$, the continuum contribution
becomes dominant. 

Strictly speaking, these observations 
only apply at high energies, as the VDM sum rule measurement at HERA was
performed at $W=82$ GeV. In absence of an independent evaluation at
lower eneriges, we are forced to extrapolate this isospin decomposition down to
lower energies. While this extrapolation is in line with the general
assumptions of the VDM, there is no guarantee that the
isospin decomposition of the photon wave function is 
energy-independent. 
To illustrate where the high-energy assumptions may break down, we note that one of the purely high energy scattering assumptions in the
derivation of the VDM sum rule is
the neglect of the real part of the forward Compton amplitude with
respect to the imaginary part. 
While this holds for the Pomeron --
whose phase is almost purely imaginary -- Reggeon exchanges contribute
to both real and imaginary part. The main contribution to
Re$\Box_{\gamma Z}$ comes from energies of $W\lesssim5$
GeV where the reggeon contribution dominates. Thus,  a re-evaluation of the
VDM sum rule at JLab energies will likely help to reduce the
theory uncertainty on the  Re$\Box_{\gamma Z}$ calculation. 

\section{Comparison to related works on dispersion $\gamma Z$ correction}

In this Section, we briefly outline the main improvements achieved in
this article with respect to our previous work, as well as the recent work
carried out by other groups. 
In  Ref. \cite{ja_chuck}, the forward dispersion relation for
$\Box_{\gamma Z}$ was derived and evaluated with the result of Re$\Box_{\gamma
  Z}\approx0.003$. 
However, that study used an over-simplified model of virtual
photoabsorption and for the sake of simplicity assumed that
$F_{1,2}^{\gamma Z}=F_{1,2}^{\gamma\gamma}$. These assumptions did not
allow for a realistic study of uncertainty of that result. 

In Ref. \cite{deltagZ_conf}, we improved on these two points: 
we employed a phenomenological model of Bosted and Christy 
that fits virtual
photoabsorption data over a large kinematic range; considered
the isospin structure of each contribution; and discussed the possible
ways to estimate uncertainty on this calculation. Although in 
\cite{deltagZ_conf} we were able to develop the general method
that we use in this manuscript, no robust theory error bar was obtained. 

Sibirtsev, Melnitchouk, Blunden and Thomas in Ref. \cite{sibirtsev} 
re-checked the findings of
Ref. \cite{ja_chuck}. That group carefully re-derived the sum rule
pointing out two errors in \cite{ja_chuck} which have been corrected here. Furthermore, the
authors of \cite{sibirtsev} proposed a model of the virtual
photoabsorption that was directly fit to the experimental
data. This allowed them to obtain an estimate for the uncertainty in the
dispersive calculation of $\Box_{\gamma Z}$ from the error bar of the
fit. To obtain the interference structure functions $F_{1,2}^{\gamma
  Z}$ from the purely electromagnetic ones $F_{1,2}^{\gamma\gamma}$, 
the authors of \cite{sibirtsev} relied on isovector dominance in the
resonance region. For the background, employed a simple scaling prescription for the
background
\be
\frac{F_2^{\gamma Z}}{F_2^{\gamma \gamma}}=\left[\frac{F_2^{\gamma Z}}{F_2^{\gamma \gamma}}\right]^{DIS}\ \ \ ,
\ee
extrapolating the isospin structure from the DIS region to low
energies. 
Ref. \cite{sibirtsev} confirmed that the dispersion correction is sizeable,
obtaining the value quoted in Table \ref{tab00}, where the error bar is due solely to fitting
the electromagnetic data. The uncertainty on the isospin
rotation of the electromagnetic data was not included. Correspondingly, the
error bar quoted in \cite{sibirtsev} only contains one part of the
total theoretical uncertainty in evaluating Re$\Box_{\gamma
  Z}$. 

Rislow and Carlson subsequently performed another computation of the dispersive contribution to 
$\Box_{\gamma Z_A}$ \cite{carlson}.
These authors again confirmed the derivation of the sum rule and
re-evaluated it obtaining the somewhat larger result quoted in Table \ref{tab00}. This was achieved by using a different
model for the resonances (the same as used in our present
work) that allows for a better fit of electromagnetic data. 
The central value of Re$\Box_{\gamma Z}$ is very close to our estimate
in forward direction, see Eq. (\ref{estimate}). 
The 
background was taken in a phenomenological form and continued into the
DIS region. The authors discussed in some detail the procedure of
isospin-rotating the resonance contributions using the constituent
quark model. 
For the background, Ref. \cite{carlson} follows to a large extent the
isovector dominance picture with a perturbation on top of that that
provides an estimate of the uncertainty. 

Both Refs. \cite{sibirtsev,carlson} argue that the uncertainty on the
dispersive calculation of Re$\Box_{\gamma Z}$ is well under control
and can be easily accomodated within the error budget of the QWEAK.
However, because they do not provide a model-independent analysis of
the isospin structure, we believe that the estimates of the error bars
obtained in those two works is unlikely to be complete.

In this manuscript, we believe we have developed the most robust and
model-independent estimate to date of the absolute size and the uncertainty of the
dispersion $\gamma Z$ correction to $A_{PV}$ in the forward limit. 
We used the most recent fit to resonance data, supplemented by two
different models of the background. We demonstrated that the two models used
in this work indeed provide a good description of the experimental
data in a very wide kinematic range of two variables $W,Q^2$.

For the $I=1/2$ resonances, we employed an
isospin rotation that is reminiscent of that for the elastic electroweak form
factors. This allows one to unambiguously relate the ratio of interference
$\gamma Z$ and the electromagnetic cross sections to combinations of transition
helicity amplitudes for the photoexcitation of a given resonance on
the proton and the neutron. We used the most recent values and
uncertainties for the latter from PDG \cite{PDG}.
The main sources of the uncertainty for the $I=1/2$ resonances is the 
neutron transition helicity amplitude of $S_{11}(1535)$, where a more 
precise extraction of the transition helicity amplitude on the
neutron would be needed.
For the $I=3/2$ resonances, the isospin rotation is straightforward.
However, we assigned a 100\% uncertainty to the contribution of
the heavy resonance state that lies close to $F_{37}(1950)$ but cannot
be reliably identified with the latter and whose isospin structure is, therefore, uncertain.

For the background we utilized two models based on the framework of vector
dominance model (VDM) that provides a prescription
for the isospin decomposition of total photoabsorption at high
energies. The VDM sum rule states an equality between the total
photoabsorption and differential cross sections for forward vector
meson production. This sum rule has been tested experimentally, albeit
only at very high energies. The precision to which this sum rule
holds  provides us with one handle for assessing for the robustness of
our isospin decomposition of the electromagnetic data. To investigate the
model dependence, we use two different models that obey the general
requirements of the VDM but originate from two kinematically distinct
regimes: Model I is a pQCD color dipole model  (what we have called the GVD/CDP approach) that is continued down 
to the real photon point by employing phenomenological input
\cite{cvetic}. Model II is a ``naive'' generalization of VDM
quoted in \cite{alwall}. 
The two models lead to similar numerical results, but within each
model the estimate of uncertainties is different. The largest contribution to our quoted theoretical
uncertainty arises from lack of knowledge of the isospin structure of the terms in these models
that are not uniquely associated with any one of the three lightest vector mesons. To be conservative, we have chosen
the largest of the corresponding uncertainties from the two models. 

Finally, we considered a phenomenological model for the intrinsic
$t$-dependence to extrapolate the forward sum rule to the experimental
kinematics. We find that the effect of such extrapolation is not
significant. However, we consider this approach to be exploratory, and an additional uncertainty on $t$-dependence may
have to be taken into account. 

To reduce the uncertainty associated with $\Box_{\gamma Z}$ to a level
below 2\%, there exist a number of avenues that could be pursued. The
most direct would be to perform measurements of the inelastic PV
asymmetries in the kinematic region that dominates the dispersion
integral: $W<5$ GeV and $Q^2\lesssim 3$ GeV$^2$. Doing so would
provide information on the electroweak structure functions
$F_{1,2}^{\gamma Z}$ 
that enter the dispersion integral for $\Box_{\gamma Z}$, thereby
mitigating 
the need for a model with which to carry out the isospin rotation. 
Additional constraints could be obtained by experimentally testing the 
VDM sum rule at the lower energies relevant to the aforementioned 
kinematics; by performing precise measurements of the electromagnetic 
neutron-to-resonance transition cross sections, thereby yielding the 
corresponding helicity amplitudes --- particularly for the
$S_{11}(1535)$; 
and identifying the isospin of the $F_{37}(1950)$ resonance. 

An alternate strategy would be to perform a measurement of $A_{PV}$ at
lower energy, given that the magnitude of, and uncertainty in,
$\Box_{\gamma Z}$ decrease monotonically with decreasing energy as
indicated in Fig. \ref{fig:deltagZ_error} and Eq. (\ref{eq:deltagZ_mami}). From the standpoint of probing physics beyond the Standard Model, a measurements of atomic PV observables for different isotopes may also be interesting. The largest atomic theory uncertainties cancel from ratios of these observables\cite{Pollock:1992mv}, and the leading sensitivity to new physics is dominated by the effects on the proton weak charge\cite{RamseyMusolf:1999qk}. To the extent that uncertainties in the neutron distributions can be constrained ({\em e.g.}, through measurements of the elastic PV asymmetry for heavy nuclei), \lq\lq isotope ratio" experiments may provide a cross-check on any inferences about new physics derived from the QWEAK measurement. Given the experimental and theoretical challenges involved in each of these efforts, an ideal program may entail a combination of the aforementioned measurements.

\acknowledgments{}
Authors are grateful to W. Melnitchouk and C. Carlson for useful
discussions and comments. 
This work was supported in part by U.S. DOE Contract
No. DE-FG02-87ER40365 and NSF Grant No. Phy-0854805
(C.J.H. and M.G.), and by U.S. DOE Contract
No. DE-FG02-08ER41531(M.J.R-M). MJRM also gratefully acknowledges support from the Wisconsin Alumni Research Foundation and the Aspen Center for Physics, where part of this work was carried out.

\begin{appendix}
\section{Isospin rotation of the resonance contributions}

In Standard Model, the $Z$ and $\gamma$ couplings to the quarks are
related by an isospin rotation, 
\beqn
J_{em}^\mu&=&q^{I=0}J_{I=0}^\mu+q^{I=1}J_{I=1}^\mu+q^sJ_s^\mu\nn\\
J_{NC_V}^\mu&=&g_V^{I=0}J_{I=0}^\mu+g_V^{I=1}J_{I=1}^\mu+g_V^sJ_s^\mu,
\eeqn
with 
\beqn
J_{I=0}^\mu&=&\frac{1}{\sqrt{2}}(\bar u\gamma^\mu u+\bar d\gamma^\mu d)\nn\\
J_{I=1}^\mu&=&\frac{1}{\sqrt{2}}(\bar u\gamma^\mu u-\bar d\gamma^\mu d)\nn\\
J_{s}^\mu&=&\bar s\gamma^\mu s
\eeqn
\indent
The e.m. charges given by $q^{I=0}=\frac{1}{3\sqrt{2}}$, 
$q^{I=1}=\frac{1}{\sqrt{2}}$, $q^s=-\frac{1}{3}$, and  
the weak charges are $g_V^{I=0}=-\frac{1}{\sqrt{2}}\frac{4}{3}s^2\theta_W$, 
$g_V^{I=1}=\frac{1}{\sqrt{2}}(2-4s^2\theta_W)$, 
$g_V^s=-1+\frac{4}{3}s^2\theta_W$.
Consequently, this isospin decomposition relates weak proton form factors
to the proton and neutron electromagnetic form factors,
\beqn
\langle p|J^\mu_{NC,V}|p\rangle=(1-4\sin^2\theta_W) \langle p|J^\mu_{em}|p\rangle
-\langle n|J^\mu_{em}|n\rangle \nn\\
\eeqn
\indent 
Above, we neglected strangeness contributions that are generally small.

A similar relation is valid for $I=\frac{1}{2}$ resonances, as well: 
\beqn
\langle X|J^\mu_{NC,V}|p\rangle=(1-4\sin^2\theta_W) \langle X|J^\mu_{em}|p\rangle
-\langle X|J^\mu_{em}|n\rangle\nn\\
\eeqn
\indent
Then, the contribution of a resonance
$R$ with isospin $1/2$ to the interference $\gamma Z$
cross section can be related 
to its contribution to the electromagnetic cross section by
\beqn
\langle p|J^\mu_{em}|R\rangle\langle R|J^\mu_{NC,V}|p\rangle&=&
(1-4s^2\theta_W) |\langle R|J^\mu_{em}|p\rangle|^2\nn\\
&-&\langle p|J^\mu_{em}|R\rangle\langle R|J^\mu_{em}|n\rangle 
\label{eq:matrel_gz_gg_def_app}
\eeqn

To proceed, we use the definition of the transition helicity amplitudes, 
\bwt
\beqn
A^{p(n)}_{R,1/2}(W^2,Q^2)=\langle R,\Lambda_R=1/2|J^\mu_{em}(\lambda_\gamma=1)|p(n),\Lambda_N=-1/2\rangle\nn\\
A^{p(n)}_{R,3/2}(W^2,Q^2)=\langle
R,\Lambda_R=3/2|J^\mu_{em}(\lambda_\gamma=1)|p(n),\Lambda_N=1/2\rangle\nn\\
S^{p(n)}_{R,1/2}(W^2,Q^2)=\langle
R,\Lambda_R=1/2|J^\mu_{em}(\lambda_\gamma=0)|p(n),\Lambda_N=1/2\rangle
\label{eq:helampl_def}
\eeqn
\ewt
where we introduced photon helicity $\lambda_\gamma=0,\pm1$, nucleon
helicity $\Lambda_N=\pm1/2$ and the helicity of the resonance $R$ that
is related to the former two as $\Lambda_R=\Lambda_N+\lambda_\gamma$.
Resonance contributions to the total cross sections $\sigma_{T,L}$ 
are related to the helicity amplitudes as 
\beqn
\sigma_T^{\gamma p(\gamma n),R}&=& \frac{2M}{M_R\Gamma_R}
\left\{|A_{R,1/2}^{p(n)}|^2+|A_{R,3/2}^{p(n)}|^2\right\},\nn\\
\sigma_{L}^{\gamma p(\gamma n),R}&=&
\frac{4M}{M_R\Gamma_R}\frac{Q^2}{q_R^2}|S_{R,1/2}^{p(n)}|^2,
\label{eq:sigmaLTgg-helampl_def_app}
\eeqn
with $M_R, \Gamma_R$ and $q_R$ the resonance mass, width and the
three-momentum of the virtual photon on the resonance position,
respectively. In the above equation, the arguments $W^2, Q^2$ of the cross
sections and helicity amplitudes were suppressed. 

We combine the definition of Eq. (\ref{eq:xi_def}) with Eqs. 
(\ref{eq:matrel_gz_gg_def_app}), (\ref{eq:helampl_def}), (\ref{eq:sigmaLTgg-helampl_def_app}), 
and finally obtain 
\beqn
\xi_{Z/\gamma}^R(Q^2)&=&(1-4s^2\theta_W)
-\frac{ A_{R,\frac{1}{2}}^p A_{R,\frac{1}{2}}^{n*}+A_{R,\frac{3}{2}}^p A_{R,\frac{3}{2}}^{n*} }
{|A_{R,\frac{1}{2}}^p|^2+|A_{R,\frac{3}{2}}^p|^2},\nn\\
\zeta_{Z/\gamma}^R(Q^2)&=&(1-4s^2\theta_W)
-\frac{ S_{R,\frac{1}{2}}^n }{S_{R,\frac{1}{2}}^p}, 
\label{eq:xigZ_helampl}
\eeqn
\indent
For spin-$\frac{1}{2}$ resonances, only $A_{1/2}^{p,n}$ pieces
contribute in the transverse ratios $\xi^R_{Z/\gamma}$. 
To a good approximation, the width and
position of a resonance can be assumed to be the same for proton and
neutron induced reactions. In this case, the $W$-dependence cancels
out in the ratio, and it is fucntion of $Q^2$ only.  

We write in general
\beqn
\xi_{Z/\gamma}^R(Q^2)&=&(1-4s^2\theta_W)-y_R\times x_R(Q^2),\nn\\
\zeta_{Z/\gamma}^R(Q^2)&=&(1-4s^2\theta_W)-\tilde y_R\times \tilde x_R(Q^2),\nn\\
\eeqn
with $y_R$ ($\tilde y_R$) the values of the ratio of the neutron and 
proton transverse (longitudinal) helicity amplitudes in Eq. (\ref{eq:helampl_def}) at
$Q^2=0$, and $x_R$ ($\tilde x_R$) the respective form factors. 
The form factors are normalized to unity at the real photon point. 
 
For the resonances of isospin $3/2$, the transition is purely
isovector, and the ratio of the cross sections is given by 
$\frac{g_V^{I=1}}{q^{I=1}}=2-4\sin^2\theta_W$, and is
$Q^2$-independent. However, for the phenomenological analyses of
the inclusive virtual photoabsorption data on the proton and neutron,
Refs. \cite{bosted,bosted2} widely used in this work, this rule does
not hold. For the $\Delta(1232)$ it holds to about 10\%. For
the $F_{37}(1950)$, the proton and neutron transition form factors show
very different behavior (monopole for the proton vs. dipole for the
neutron). Furthermore, the unnaturally mild monopole form factor 
raises a question of whether this contribution should be
considered as part of the background where monopole form factors arise
naturally in the VDM picture. 
Correspondingly, rather than operate with a form factor $x_R(Q^2$ for
the two isospin-$3/2$ resonances we will assign an uncertainty to the
ratios $y_R^{I=3/2}=-1$: 10\% for the $\Delta(1232)$ and 100\% for the
$F_{37}(1950)$, and use $x_R(Q^2)=1$ for both. 

We next turn to the form factors $x_R(Q^2)$ of the isospin-1/2
resonances. To estimate these, 
one needs the $Q^2$-dependence of the transition helicity amplitudes
for the excitation of these resonances. Unfortunately, the
phenomenological fits of Refs. \cite{bosted,bosted2} do not provide us
with this information: they only give us
$|A_{R,1/2}^{p(n)}|^2+|A_{R,3/2}^{p(n)}|^2$ and
$|S_{R,1/2}^{p(n)}|^2$. Instead, we need e.g., 
$A_{R,\frac{1}{2}}^p A_{R,\frac{1}{2}}^{n*}+A_{R,\frac{3}{2}}^p
A_{R,\frac{3}{2}}^{n*}$.
For spin-1/2 resonances, only $A_{1/2}$'s contribute. Then, one has 
for the second terms in Eq. (\ref{eq:xigZ_helampl}) 
\beqn
\frac{ A_{R,\frac{1}{2}}^p
  A_{R,\frac{1}{2}}^{n*}}{|A_{R,\frac{1}{2}}^p|^2}&=&
\pm\sqrt{\frac{\sigma_T^{\gamma n,R}}{\sigma_T^{\gamma p,R}}}\,,\nn\\
\frac{ S_{R,\frac{1}{2}}^p
  S_{R,\frac{1}{2}}^{n*}}{|S_{R,\frac{1}{2}}^p|^2}&=&
\pm\sqrt{\frac{\sigma_L^{\gamma n,R}}{\sigma_L^{\gamma p,R}}}\,,
\label{eq:xR_spin1/2}
\eeqn
and the only missing piece above is the relative sign
of the proton and neutron helicity amplitudes. This sign is well defined and
can be taken, for instance, from the PDG or from quark model
\cite{koniuk}. For spin-3/2 resonances $D_{13}(1520)$ and
$F_{15}(1680)$, the information provided by
Refs. \cite{bosted,bosted2} is not sufficient to determine respective
$x_R(Q^2)$. We can only approximately estimate those by noticing that
for these resonances, the PDG suggests that the $p\to N^*$ transition
is completely dominated by the $A_{3/2}$ helicity amplitude
\cite{PDG}. Then, we can adapt the same logic as for the
spin-1/2 resonances, by substituting $A_{3/2}$'s in place of $A_{1/2}$'s
in Eq. (\ref{eq:xR_spin1/2}). 

However, this procedure cannot be considered reliable since such
``extracted'' form factors $x_R(Q^2)$ will contain a model dependence
that is very hard to estimate. Instead, we will use the following
reasoning. We verified that with the approximations described above,
the results of Refs. \cite{bosted,bosted2} lead to the form factors
$x_R(Q^2)$ that differ from 1 by at most 10-20\% for values of
$Q^2\leq0.6-0.8$ GeV$^2$ for all five isospin-1/2 resonances. 
At the same time, the PDG quotes the errors
for the helicity amplitudes \cite{PDG} for the excitation of those resonances
that are conservative enough to accomodate these 10-20\% discrepancy. 
Indeed, the PDG values represent an average over world data and over
various analyses, therefore the errors that they quote contain not
only the statistical and systematic error of each experiment, but
also the systematic error due to model dependence of those
analyses.
This means that at low values of $Q^2$, the error
introduced if setting $x_R(Q^2)\approx x_R(0)=1$ is reasonably small
as compared to the error in the respective $y_R$. While at larger
values of $Q^2$ this is not the case any longer, due to resonance form
factors the impact
of these values of $Q^2$ on the dispersion correction $\Box_{\gamma
  Z}$ is small. From the discussion of the results, 
we see that 
\begin{itemize}
\item{} the resonance contribution is dominated by the
$\Delta(1232)$ for which the issue of the uncertainty in $x_R(Q^2)$
is controlled within 10\%, as discussed earlier;
\item{} the overall uncertainty on the resonance contribution is
  dominated by that due to the problem of the identification of the high lying resonance
in the analysis of Bosted and Christy with the $F_{37}(1950)$;
\item{} the total uncertainty in the dispersion correction $\Box_{\gamma
  Z}$ is dominated by the uncertainty due to the background
contribution. Then, even doubling the uncertainty in the contribution
of the $S_{11}(1535)$ due to $x_R(Q^2)$ will not significatly 
change our overall conclusions.
\end{itemize}
This allows us to set all $x_R(Q^2)=1$ for all seven
resonances considered here (including the isospin-3/2 resonances
discussed earlier). The error introduced by this approximation is
safely covered by using the conservative PDG errors for the resonance
helicity amplitudes.  

Finally, we discuss the ratios of the longitudinal cross sections
$\zeta_{Z/\gamma}^R$. In Ref. \cite{bosted2}, it was shown that the
hypothesis that the ratio of the resonance contributions to 
longitudinal and the transverse cross
sections for the proton and for the neutron target are equal, is well supported
by the experimental data. Although this conclusion is model-dependent,
as well, the general impact of the longitudinal cross section on the 
$\Box_{\gamma Z}$ was found to be very small. This allows us to use
the assumption of Ref. \cite{bosted2} here and set 
$\zeta_{Z/\gamma}^R =\xi_{Z/\gamma}^R$ for all seven resonances. 
As a result, we arrive at Eq. (\ref{F12_rescale}) with
$\xi_{Z/\gamma}^R=1-4\sin^2\theta_W-y_R$ where the values of $y_R$ are
listed in Table \ref{tab}.

\end{appendix}

\end{document}